\documentclass[preprint,5p,12pt,twocolumn]{article}

\usepackage{amssymb}
\usepackage{subfig}
\usepackage{algorithm,algorithmic}
\usepackage{graphicx}
\usepackage{caption}
\usepackage{xspace}
\usepackage{multirow}
\usepackage{rotating}
 \usepackage{url}
\usepackage{gensymb}
\usepackage[dvipsnames,table]{xcolor}
\usepackage[top=2cm, bottom=2.2cm, left=1.5cm, right=1.5cm]{geometry}
\usepackage{setspace}
\usepackage{morefloats}
\usepackage[keeplastbox]{flushend}

\usepackage{multirow}
\usepackage{url}
\usepackage[hyphenbreaks]{breakurl}
\usepackage{amsmath}

\usepackage[affil-it]{authblk}
\usepackage[hidelinks]{hyperref}
\hypersetup{
	colorlinks = true,
	citecolor = blue,
}
\usepackage{pdfpages}
\usepackage{eso-pic}
\definecolor{backcream}{HTML}{FFF0C8}

\AddToShipoutPictureFG*{
	\AtPageUpperLeft{%
		\raisebox{-50pt}{\makebox[\paperwidth]{\noindent\fcolorbox{red}{backcream}{\begin{minipage}{18.5cm}\centering
					{\color{red}\textbf{This manuscript is accepted in IEEE Transactions on Neural Networks and Learning Systems.} \textbf{Please cite it as:}}\\
					Mozafari, M., Kheradpisheh, S. R., Masquelier, T., Nowzari-Dalini, A., \& Ganjtabesh, M. (2018). First-Spike-Based Visual Categorization Using Reward-Modulated STDP. IEEE Transactions on Neural Networks and Learning Systems (\url{https://doi.org/10.1109/TNNLS.2018.2826721}).
				\end{minipage}}}}%
			}
		}

\begin{document}
\title{\LARGE \textbf{First-spike based visual categorization using\\reward-modulated STDP}}

\author{Milad Mozafari$ ^{1,2}$}
\author{Saeed Reza Kheradpisheh$ ^{1,2}$}
\author{Timoth\'ee Masquelier$ ^{3}$}
\author{Abbas Nowzari-Dalini$ ^{1}$}
\author{Mohammad Ganjtabesh$ ^{1,2,}$\footnote{Corresponding author. \\ \hspace*{0.5cm} Email addresses:\\ \hspace*{1cm} milad.mozafari@ut.ac.ir (MM), \\ \hspace*{1cm} kheradpisheh@ut.ac.ir (SRK), \\ \hspace*{1cm} timothee.masquelier@cnrs.fr (TM), \\ \hspace*{1cm}  nowzari@ut.ac.ir (AND), \\ \hspace*{1cm} mgtabesh@ut.ac.ir (MG).}}
\affil{\footnotesize $ ^{1} $ Department of Computer Science, School of Mathematics, Statistics, and Computer Science, University of Tehran, Tehran, Iran}
\affil{\footnotesize $ ^{2} $ School of Biological Sciences, Institute for Research in Fundamental Sciences (IPM) , Tehran, Iran}
\affil{\footnotesize $ ^{3} $ CerCo UMR 5549, CNRS — Universit\'e Toulouse 3, France}

\date{}

\maketitle
\begin{abstract}
Reinforcement learning (RL) has recently regained popularity, with major achievements such as beating the European game of Go champion. Here, for the first time, we show that RL can be used efficiently to train a spiking neural network (SNN) to perform object recognition in natural images without using an external classifier. We used a feedforward convolutional SNN and a temporal coding scheme where the most strongly activated neurons fire first, while less activated ones fire later, or not at all. In the highest layers, each neuron was assigned to an object category, and it was assumed that the stimulus category was the category of the first neuron to fire. If this assumption was correct, the neuron was rewarded, i.e. spike-timing-dependent plasticity (STDP) was applied, which reinforced the neuron's selectivity. Otherwise, anti-STDP was applied, which encouraged the neuron to learn something else. As demonstrated on various image datasets (Caltech, ETH-80, and NORB), this reward modulated STDP (R-STDP) approach extracted particularly discriminative visual features, whereas classic unsupervised STDP extracts any feature that consistently repeats. As a result, R-STDP outperformed STDP on these datasets. Furthermore, R-STDP is suitable for online learning, and can adapt to drastic changes such as label permutations. Finally, it is worth mentioning that both feature extraction and classification were done with spikes, using at most one spike per neuron. Thus the network is hardware friendly and energy efficient.
	
	\vspace{0.3cm}
\end{abstract}

\section{Introduction}\label{intro}
Neurons in the brain are connected by synapses that can be strengthened or weakened over time. The neural mechanisms behind long-term synaptic plasticity, which is crucial for learning, have been under investigation for many years. Spike-timing-dependent plasticity (STDP) is an unsupervised form of synaptic plasticity, observed in different brain areas~\cite{gerstner1996neuronal,markram1997regulation,bi1998synaptic,sjostrom2001rate}, in particular in the visual cortex~\cite{meliza2006receptive, huang2014associative, guo2017timing}. STDP works by considering the time difference between pre- and post-synaptic spikes. According to this rule, if the pre-synaptic neuron fires earlier (later) than the post-synaptic one, the synapse is strengthened (weakened). Studies have shown that STDP results in coincidence detectors, by which a neuron gets selective to a frequent input spike pattern leading to an action potential whenever the pattern is presented~\cite{masquelier2008spike, gilson2011stdp, brette2012computing, masquelier2017stdp}. STDP works well in finding statistically frequent features, however, as any unsupervised learning algorithm, it faces with difficulties in detecting rare but diagnostic features for important functionalities such as decision-making.

Several studies suggest that the brain's reward system plays a vital role in decision-making and forming behaviors. This is also known as reinforcement learning (RL), by which the learner is encouraged to repeat rewarding behaviors and avoid those leading to punishments~\cite{sutton1998introduction, dayan2002reward, daw2006computational, niv2009reinforcement, lee2012neural,steinberg2013causal, schultz2015neuronal}.  It is found that dopamine, as a neuromodulator, is one of the important chemical substances involved in the reward system~\cite{schultz2002getting}, where its release is proportional to the expected future reward~\cite{steinberg2013causal, schultz1998predictive, glimcher2011understanding}. It is also shown that dopamine, as well as some other neuromodulators influences the synaptic plasticity, such as changing the polarity~\cite{seol2007neuromodulators} or adjusting the time window of STDP~\cite{gu2002neuromodulatory,reynolds2002dopamine,zhang2009gain, marder2012neuromodulation, nadim2014neuromodulation}.

One of the well-studied ideas to model the role of the reward system is to modulate or even reverse the weight change determined by STDP, which is called reward-modulated STDP (R-STDP)~\cite{fremaux2016neuromodulated}. R-STDP stores the trace of synapses that are eligible for STDP and applies the modulated weight changes at the time of receiving a modulatory signal; a reward or punishment (negative reward).

In 2007, Izhikevich~\cite{izhikevich2007solving} proposed a R-STDP rule to solve the distal reward problem, where the reward is not immediately received. He solved the problem using a decaying eligibility trace, by which the recent activities are considered to be more important. He showed that his model can solve both classical and instrumental conditionings~\cite{pavlov2003conditioned, thorndike1898review}. In the same year, Farries and Fairhall~\cite{farries2007reinforcement} employed R-STDP to train neurons for generating particular spike patterns. They measured the difference between the output and target spike trains to compute the value of the reward. Also, Florian~\cite{florian2007reinforcement} showed that R-STDP is able to solve the XOR task by either rate or temporal input coding and learning a target firing rate. A year later, Legenstein et al.~\cite{legenstein2008learning} investigated conditions, under which R-STDP achieves a desired learning effect. They demonstrated the advantages of R-STDP by theoretical analysis, as well as practical applications to biofeedbacks and a two-class isolated spoken digit recognition task. Vasilaki et al.~\cite{vasilaki2009spike} examined the idea of R-STDP on problems with continuous space. They showed that their model is able to solve the Morris water maze quite fast, while the standard policy gradient rule failed. Investigating capabilities of R-STDP continued by research from Fr\'emaux et al.~\cite{fremaux2010functional}, in which conditions for a successful learning is theoretically discussed. They showed that a prediction of the expected reward is necessary for R-STDP to learn multiple tasks simultaneously. Studying the RL mechanism in the brain has gathered attentions in recent years, and researchers try to solve more practical tasks by reward-modulated synaptic plasticity~\cite{friedrich2011spatio, fremaux2013reinforcement, hoerzer2014emergence}.

Visual object recognition is a sophisticated task, at which humans are expert. This task requires both feature extraction, that is done by the brain's visual cortex, and decision-making on the category of the object, for which higher brain areas are involved. Spiking neural networks (SNNs) have been widely used in computational object recognition models. In terms of network architecture, there are several models with shallow~\cite{masquelier2007unsupervised, brader2007learning, querlioz2013immunity, yu2013rapid}, deep~\cite{lee2016training, o2016deep, kheradpisheh2017stdp}, recurrent~\cite{thiele2017wake}, fully connected~\cite{diehl2015unsupervised}, and convolutional structures~\cite{masquelier2007unsupervised,kheradpisheh2017stdp, cao2015spiking, tavanaei2016bio}. Some use rate-based coding ~\cite{merolla2011digital,hussain2014improved,oconnor2014realtime}, while others use the temporal coding~\cite{masquelier2007unsupervised, yu2013rapid, kheradpisheh2017stdp,diehl2015unsupervised, beyeler2013categorization}. Various kinds of learning techniques are also applied to SNNs, from backpropagation~\cite{cao2015spiking,diehl2015fast}, tempotron~\cite{yu2013rapid,zhao2015feedforward}, and other supervised techniques~\cite{hussain2014improved,oconnor2014realtime, ponulak2010supervised,neftci2012event}, to unsupervised STDP and STDP-variants~\cite{querlioz2013immunity, diehl2015unsupervised, tavanaei2016acquisition}. Although STDP-enabled networks provide a more biological plausible means of visual feature extraction, they need an external readout, e.g. support vector machines~\cite{kheradpisheh2017stdp, kheradpisheh2016bio}, to classify input stimuli. Additionally, STDP tends to extract frequent features which are not necessarily suitable for the desired task. In this research, we present a hierarchical SNN equipped with R-STDP to solve the visual object recognition in natural images, without using any external classifier. Instead, we put class-specific neurons that are reinforced to fire as early as possible if their target stimulus is presented to the network. Thus, the input stimuli are classified solely based on the first-spike latencies in a fast and biologically plausible way. R-STDP enables our network to find task-specific diagnostic features, therefore, decreases the computational cost of the final recognition system.

Our network is based on Masquelier and Thorpe’s model~\cite{masquelier2007unsupervised} with four layers. The first layer of the network converts the input image into spike latencies based on the saliency of its oriented edges. This spike train goes under a local pooling operation in the second layer. The third layer of the network includes several grids of integrate-and-fire neurons that combine the received information of oriented edges and extract complex features. This is the only trainable layer in our network which employs R-STDP for synaptic plasticity. The signal (reward/punishment) for modulation of synaptic plasticity is provided by the fourth layer, in which the decision of the network is made. Our network only uses the earliest spike emitted by the neurons in the third layer to make a decision, without using any external classifier. If its decision is correct (incorrect) a global reward (punishment) signal is generated. Besides, in order to increase the computational efficiency, each cell in the network is allowed to spike only once per image. The motivation for at most one spike per neuron is not only computational efficiency, it is also biological realism~\cite{thorpe1989biological, vanrullen2002surfing}. Decision-making without any classifiers with at most one spike per neuron, turns the proposed method into a well-suited candidate for the hardware implementation.

We performed two toy experiments to illustrate the abilities of R-STDP. We showed that the network employing R-STDP finds informative features using fewer computational resources than STDP. We also showed that R-STDP can change the behavior of a neuron, if needed, by encouraging it to unlearn what it has learned before. Thus, reusing a computational resource that is no longer useful. Moreover, we evaluated the proposed network on object recognition in natural images, using three different benchmarks, that are Caltech face/motorbike (two classes), ETH-80 (eight classes), and NORB (five classes). The results of the experiments demonstrate the advantage of employing R-STDP over STDP in finding task-specific discriminative features. Our network reached the performances (recognition accuracies) of $98.9\%$ on Caltech face/motorbike, $89.5\%$ on ETH-80, and $88.4\%$ on NORB datasets.

The rest of this paper is organized as follows: A precise description of the proposed network is provided in Section $2$. Then, in Section $3$, the results of the experiments are presented. Finally, in Section $4$, the proposed network is discussed from different points of view and the possible future works are highlighted.

\section{Materials and Methods}\label{matmet}
In this section, we first describe the structure of the proposed network and the functionality of each layer. We then explain R-STDP, by which the neurons achieve reinforced selectivity to a specific group of input stimuli. Finally, we give a detailed description on the classification strategy that is used to evaluate the network's performance.

\subsection{Overall Structure}
\begin{figure}[t]
	\begin{center}
		\includegraphics[width = 8.5cm]{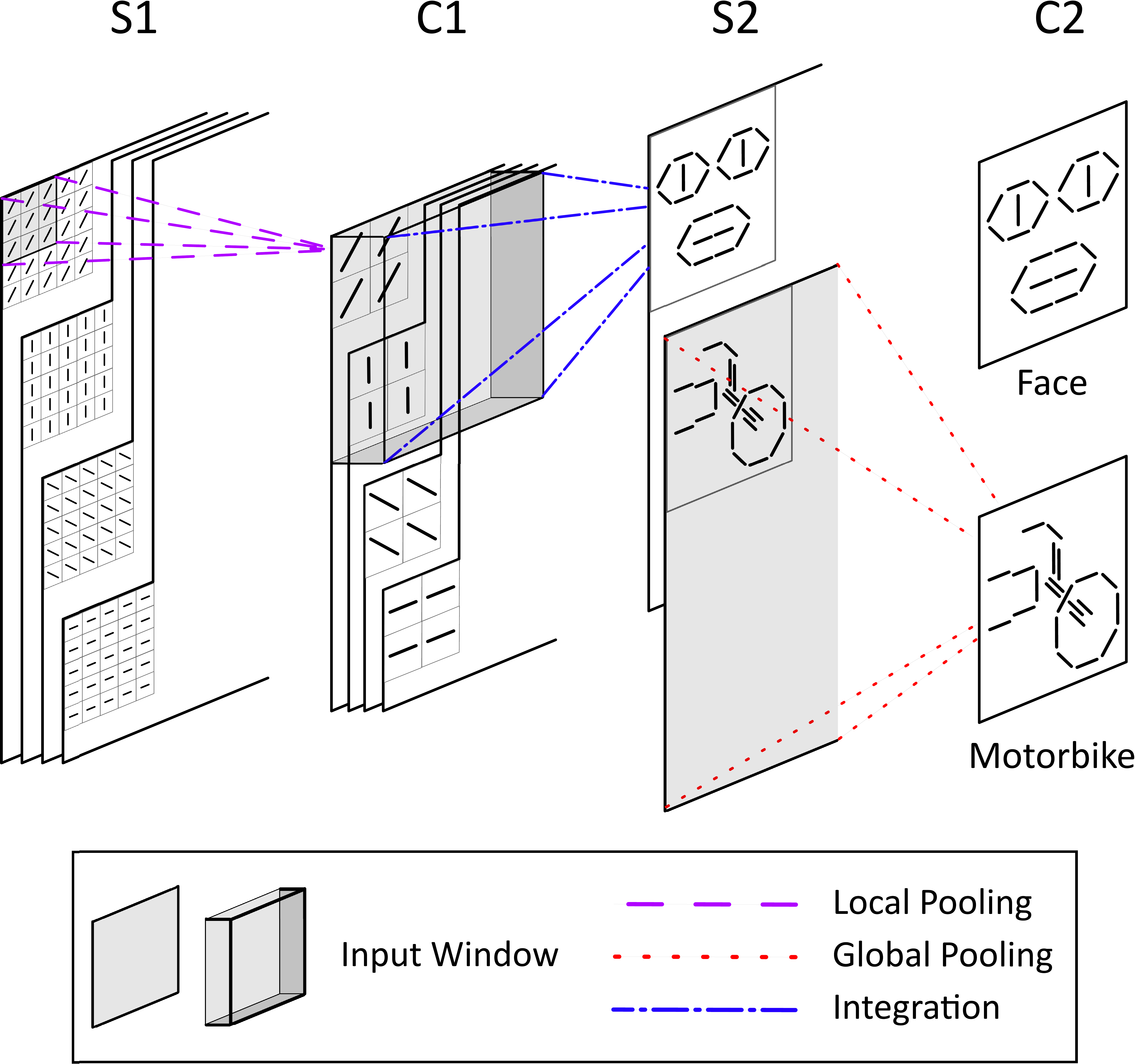}
		\caption{Overall structure of the proposed network with four retinotopically organized layers. The first layer ($S1$) extracts oriented edges from the input image by applying Gabor filters. A local max-pooling operation is applied by the cells in the subsequent layer ($C1$) to gain some degrees of position invariance. From here, spikes are propagated by the latencies that are inversely proportional to the maximum values. These spikes are the inputs for the IF neurons in the layer $S2$ that are equipped with the R-STDP learning rule. These neurons are encouraged/punished to learn/unlearn complex features. The activity of $S2$ neurons are used by $C2$ neurons for decision-making. These neurons are associated with class labels and the decision is made based on the neuron with the earliest spike.}
		\label{fig:network_structure}
	\end{center}
\end{figure}

Similar to Masquelier and Thorpe’s model~\cite{masquelier2007unsupervised}, our network consists of two simple and two complex layers, that are alternately arranged in a feed-forward manner (see Fig.~\ref{fig:network_structure}).

The first layer of the network ($S1$) is a simple layer whose cells detect oriented edges in the input image. These cells emit a spike with a latency that is inversely proportional to the saliency of the edge. After $S1$, there is a complex layer ($C1$), which introduces some degrees of position invariance by applying local pooling operation. A $C1$ neuron propagates the earliest spike in its input window.

The second simple layer ($S2$) is made of integrate-and-fire (IF) neurons. A neuron in this layer, that detects a complex feature, receives its inputs from $C1$ neurons and generates a spike once its membrane potential reaches the threshold. For synaptic plasticity, we use a learning rule based on three factors: (1) pre-synaptic spike time, (2) post-synaptic spike time, and (3) a reward/punishment signal. This kind of synaptic plasticity provides the ability to control the behavior of the neurons in terms of their selectivity to input patterns.

The second complex layer ($C2$) of our network is the decision-making layer. Each neuron in this layer is assigned to a category and performs a global pooling operation over $S2$ neurons in a particular grid. Using a rank-order decoding scheme, the neuron which fires first, indicates the network's decision about the input image. According to the decision made by the network, a reward/punishment signal is then generated, which drives in the synaptic plasticity of $S2$ neurons.

Implementation of the network is mainly done with C\# and the code is available on ModelDB\footnote{\url{https://senselab.med.yale.edu/ModelDB/}}.

\subsection{Layer $S1$}
The goal of this layer is to extract oriented edges from the gray scaled input image and turn them into spike latencies. To this end, the input image is convolved with Gabor filters of four different orientations. Thus, this layer includes four feature maps, each representing the saliency of edges in a particular preferred orientation.

Let $I$ be the grayscaled input image and $G(\theta)$ represent a Gabor filter (convolution kernel) with window size $5\times 5$, wavelength $2.5$, effective width $2$, and orientation $\theta$. Then, the $l$th feature map of layer $S1$ is generated using the following equations:
\begin{equation}
\begin{split}
S_{1}^l &= |I\otimes G(\theta_{l})|,\\
\theta_{l} &= \frac{(l - 1)\times \pi}{4} +‌\frac{\pi}{8},
\end{split}
\end{equation}
where $\otimes$ is the convolution operator and $l \in \lbrace 1, 2, 3, 4 \rbrace$. In order to introduce invariance to image negative operation, the absolute value of the convolution is used. Also, since vertical and horizontal edges are very common in natural images, a $\frac{\pi}{8}$ offset is applied to relax this bias~\cite{masquelier2007unsupervised}.

For each of the feature maps (orientations), we put a 2D grid of the same size containing dummy neurons to propagate spikes. Using an intensity-to-latency encoding scheme, the obtained feature maps are converted to the spike latencies that are inversely proportional to the saliency of edges. In other words, the more salient the edge, the earlier the corresponding spike is propagated.

We implemented the proposed network in an event-based manner, where the spikes are sorted by their latencies in ascending order and propagated sequentially (i.e. the first spike is propagated in time step $t = 1$, the second one in $t = 2$, and so on).

\subsection{Layer $C1$}
Our first complex layer is a local pooling layer over the spikes coming from layer $S1$. Here, there are four 2D neuronal grids corresponding to each of the orientations. Each $C1$ neuron performs a local pooling operation over a window of size $\omega_{c1}\times \omega_{c1}$ and stride $r_{c1}$ (here we set $r_{c1} = \omega_{c1} - 1$) on $S1$ neurons in a particular grid, after which, it emits a spike immediately after receiving its earliest input spike. This pooling operation decreases the redundancy of layer $S1$, and shrinks the number of required neurons, which consequently increases the computational efficiency. It also adds a local invariance to the position of oriented edges.

Let $\mathcal{P}_{c1}(i)$ be the set of all pre-synaptic neurons of the $i$th neuron in layer $C1$. Then, the firing time of this neuron is computed as follows:

\begin{equation}
t_{c1}^f(i) = \min_{j \in \mathcal{P}_{c1}(i)} \{t_{s1}^f(j)\},
\end{equation}
where $t_{s1}^f(j)$ denote the firing time of the $j$th neuron in $\mathcal{P}_{c1}(i)$.

Additionally, two kinds of lateral inhibition mechanisms are employed, which help the network to propagate more salient information. If a neuron located at position $(x,y)$ of the $i$th grid (orientation) fires, (1) the other neurons at the same position, but in other grids are prevented from firing, and (2) the latencies of the nearby neurons in the same grid are increased by a factor relative to their mutual Euclidean distance. In our experiments, inhibition is done for distances from $1$ to $5$ pixel(s) (floating-point distances are truncated to integer values) with inhibition factors $15\%, 12\%, 10\%, 7\%,$ and $5\%$, respectively.

\subsection{Layer $S2$}
This layer combines the incoming information about oriented edges and turns them into meaningful complex features. Here, there are $n$ 2D grids of IF neurons with the threshold $\mathcal{T}$. Each neuron receives its inputs from a $\omega_{s2}\times \omega_{s2}\times 4$ window of $C1$ neurons through plastic synapses. A weight sharing mechanism is also applied to the neurons belonging to the same grid. This mechanism provides the ability of detecting a particular feature over the entire spatial positions. To be precise, let $\mathcal{P}_{s2}(i)$ be the set of all pre-synaptic neurons corresponding to the $i$th neuron. Then, the membrane potential of this neuron at time step $t$ is updated by the following equation:
\begin{equation}
v_i(t) = v_{i}(t - 1) + \sum_{j \in \mathcal{P}_{s2}(i)}^{} {W_{ij}\times \delta\left(t-t^f_{c1}(j)\right)},
\end{equation}
where $W_{ij}$ denotes the synaptic weight, $\delta$ is the Kronecker delta function, and $t^f_{c1}(j)$ is the firing time of the $j$th cell in layer $C1$. For each input image, a neuron in $S2$ fires if its membrane potential reaches the threshold $\mathcal{T}$. Also, these neurons have no leakage and are allowed to fire at most once while an image is being presented.

As the neurons fire, their synaptic weights - the feature they are detecting - are being updated based on the order of pre- and post-synaptic spikes, as well as a reward/punishment signal (see section reward-modulated STDP). This signal is derived from the activity of the next layer, that indicates the network's decision. Besides, initial weights of the synapses are
randomly generated, with mean $0.8$ and standard deviation $0.05$. Note that choosing small or midrange values for mean results in inactive, thus untrained, neurons. Moreover, large values for variance increase the impact of network's initial state. Accordingly, high mean value with small variance is a suitable choice~\cite{kheradpisheh2017stdp}.

\subsection{Layer $C2$}
This layer contains exactly $n$ neurons, each is assigned to one of the $S2$ neuronal grids. A $C2$ neuron only propagates the first spike that is received from its corresponding neuronal grid. To put it differently, let $\mathcal{P}_{c2}(i)$ define the set of $S2$ neurons in the $i$th neuronal grid (for $i \in \{1,2,...,n\}$). Then, the firing time of the $i$th $C2$ neuron is computed as follows:

\begin{equation}
t_{c2}^f(i) = \min_{j \in \mathcal{P}_{c2}(i)} \{t_{s2}^f(j)\},
\end{equation}
where $t_{s2}^f(j)$ denote the firing time of the $j$th neuron in layer $S2$.

As mentioned before, the activity of $C2$ neurons indicates the decision of the network. To this end, we divide $C2$ neurons into several groups and assign each group to a particular category of input stimuli. Then, the network's decision on the category of the input stimulus is assumed to be the one whose group propagates the earliest spike among other $C2$ groups.

Assume that there are $m$ distinct categories for the input stimuli, labeled from $1$ to $m$, and $n$ neuronal grids in layer $S2$. Accordingly, there are exactly $n$ neurons in layer $C2$, that are divided into $m$ groups. Let $g: \{1,2,...,n\} \mapsto \{1,2,...,m\}$ denote a function that returns the group's index of a $C2$ neuron, and let $t^f_{c2}(i)$ denote the firing time of the $i$th neuron in layer $C2$. Then, the network's decision $\mathcal{D}$ is made by

\begin{equation}\label{eq:decision}
\begin{split}
\mathcal{F} &= \arg\min_{i} \lbrace t^f_{c2}(i) | 1\leq i \leq n \rbrace,\\
\mathcal{D} &= g\left(\mathcal{F}\right),
\end{split}
\end{equation}
where $\mathcal{F}$ is the index of a $C2$ neuron which fires first. The network receives reward (punishment) if its decision matches (does not match) the correct category of the input stimulus. If none of the $C2$ neurons fire, no reward/punishment signal is generated, thus, no weight-change is applied. Moreover, if more than one neuron fire early (with the minimum spike time), the one with the minimum index ($i$) is selected.

\subsection{Reward-Modulated STDP (R-STDP)}\label{sec:rstdp}
We propose a reinforcement learning mechanism to update the pre-synaptic weights of $S2$ neurons. Here, the magnitude of weight change is modulated by a reward/punishment signal, which is received according to the correctness/incorrectness of the network's decision. We also applied a one-winner-takes-all learning competition among the $S2$ neurons, by which the one with the earliest spike is the winner and the only one which updates its synaptic weights. Note that this neuron is the one determining the network's decision.

To formulate our R-STDP learning rule, if a reward signal is received, then
\footnotesize
\begin{equation}
\hspace{-0.07cm}\Delta W_{ij}\!=\!
\begin{cases}
a_r^+\!\times\!W_{ij}\!\times\!\left(1\!-\!W_{ij}\right) & \mbox{if} \ t^f_{c1}(j) - t^f_{s2}(i) \leq 0,\vspace{0.4cm}\\
a_r^-\!\times\!W_{ij}\!\times\!\left(1\!-\!W_{ij}\right) &
\begin{aligned}
\mbox{if} \ & t^f_{c1}(j) - t^f_{s2}(i) > 0,\\ &\mbox{\hspace{-0.5cm}or the $j$th cell is silent,}
\end{aligned}
\end{cases}
\end{equation}
\normalsize
and in case of receiving a punishment signal, we have
\footnotesize
\begin{equation}
\hspace{-0.07cm}\Delta W_{ij}\!=\!
\begin{cases}
a_p^+\!\times\!W_{ij}\!\times\!\left(1\!-\!W_{ij}\right) &
\begin{aligned}
\mbox{if} \ & t^f_{c1}(j)\!-\!t^f_{s2}(i) > 0,\\ &\mbox{\hspace{-0.5cm}or the $j$th cell is silent,}
\end{aligned}\vspace{0.4cm}\\
a_p^-\!\times\!W_{ij}\!\times\!\left(1\!-\!W_{ij}\right) & \mbox{if} \ t^f_{c1}(j) - t^f_{s2}(i) \leq 0,
\end{cases}
\end{equation}
\normalsize
where $i$ and $j$ refer to the post- and pre-synaptic cells, respectively, $\Delta W_{ij}$ is the amount of weight change for the synapse connecting the two neurons, and $a_r^+$, $a_r^-$, $a_p^+$, and $a_p^-$ scale the magnitude of weight change. Furthermore, to specify the direction of weight change, we set $a^+_r, a^+_p > 0$ and $a^-_r, a^-_p < 0$. Here, our learning rule does not take into account the exact spike time difference and uses an infinite time window. According to this learning rule, the punishment signal reverses the polarity of STDP (a.k.a anti-STDP). In other words, it swaps long-term-depression (LTD) with long-term-potentiation (LTP), which is done to conduct the effect of aversion (avoid repeating a bad behavior), and $a^+_p$ is there to encourage the neuron to learn something else.

\subsection{Overfitting Avoidance}
In reinforcement learning problems, there is a chance of being trapped into  local optima or overfitting to acquiring the maximum possible reward over the training examples. In order to help the network, exploring other possible solutions that are more general to cover both seen and unseen examples, we apply two additional mechanisms during the training phase. These techniques are only used for object recognition tasks.

\subsubsection{Adaptive learning rate}
Since the initial weights of the neurons are randomly set, the number of misclassified samples is relatively high at the beginning of the training phase (i.e. the performance is at the chance level). As training trials go on, the ratio of correctly classified samples to the misclassified ones increases. In the case of high rate of misclassification, the network receives more punishment signals, which rapidly weakens synaptic weights and generates dead or highly selective neurons that cover a small number of inputs. Similarly, when the rate of correct classification gets higher, the rate of reward acquisition increases as well. In this case, the network prefers to exclude misclassified samples by getting more and more selective to correct ones and remain silent for the others. In either cases, the overfitting happens due to the unbalanced impact of reward and punishment.

To tackle this problem, we multiply an adjustment factor to the amount of weight modification, by which the impact of correct and incorrect training samples is balanced over the trials. Assume that the network sees all of the training samples on each training iteration and let $N_{hit}$ and $N_{miss}$ denote the number of samples that are classified correctly and incorrectly in the last training iteration, respectively. If $N$ is the number of all training samples, then, the weight changes for the current training trial are modified as follows:
\footnotesize
\begin{equation}
W_{ij} = W_{ij} + 
\begin{cases}
\left(\frac{N_{miss}}{N}\right)\Delta W_{ij} & \mbox{if a reward is received},\vspace{0.3cm}\\
\left(\frac{N_{hit}}{N}\right)\Delta W_{ij} & \mbox{otherwise}.
\end{cases}
\end{equation}
\normalsize
Note that $N_{hit} + N_{miss} \leq N$, since there may be some samples for which none of the $S2$ neurons is active.

\subsubsection{Dropout}
In a reinforcement learning scenario, the goal of the learner is to maximize the expected value of reward acquisition. In our case, since the network only sees the training samples, it may find a few number of features that are sufficient to correctly classify almost all of the training samples. This issue appears to cause severe overfitting in face of complex problems and the network prefers to leave some of the neurons untrained. These neurons decrease the hit rate of the network over the testing samples, as they blindly fire for almost all of the stimuli.

Here, we employ the dropout technique~\cite{krizhevsky2012imagenet}, which causes a $C2$ neuron to be temporary turned off with the probability of $p_{drop}$. This technique gives rise to the overall involvement rate of the neurons, which in turn, not only increases the chance of finding more discriminative features, but also decreases the rate of blind firings (see Supplementary Materials: Dropout).

\subsection{Classification}
As mentioned before, the activity of the last layer, particularly the earliest spike in layer $C2$, is the only information that our network uses to make its final decision on the input stimuli. This way, we do not need external classifiers and increase the biological plausibility of the network at the same time.

To setup the network for a classification task with $m$ categories, we put $n = k \times m$ neuronal grids in layer $S2$, where $k$ is the number of features associated to each category. Then, we assign each $C2$ neurons to a category by the association function $g: \{1,2, ..., n\} \mapsto \{1, 2, ..., m\}$ defined as follows:
\begin{equation}
g(i) = \lfloor (i - 1)/k \rfloor + 1.
\end{equation}
Then, the network uses equation (\ref{eq:decision}) to classify the input stimuli. During the training phase, each network's decision is compared to the label of stimulus and a reward (punishment) signal is generated, if the decision matches (mismatches) the label.

\subsection{Comparison of R-STDP and STDP}
In object recognition tasks, we make a comparison between our model, SNN with R-STDP, and the one that uses STDP. To this end, we first train the network using STDP and let the network extract features in an unsupervised manner. Next, we compute three kinds of feature vectors of length $n$ from layer $S2$:

\begin{itemize}
	\item \textbf{The first-spike vector}. This is a binary vector, in which all the values are zeros, except the one corresponding to the neuronal grid with earliest spike.
	\item \textbf{The spike-count vector}. This vector saves the total number of spikes emitted by neurons in each grid.
	\item \textbf{The potential vector}. This vector contains the maximum membrane potential among the neurons in each grid, by ignoring the threshold.
\end{itemize}

After extracting feature vectors for both training and testing sets, K-nearest neighbors (KNN) and support vector machine (SVM) classifiers are used to evaluate the performance of the network. Moreover, the learning strategy and the STDP formula is the same as~\cite{masquelier2007unsupervised}, and to make a fair comparison, we use the same values for parameters in both models. The only parameters that are explored for the STDP are the magnitudes of LTP and LTD.

\section{Results}\label{result}
To evaluate the proposed network and learning strategy, we performed two types of experiments. First, we used a series of hand-made problems to show the superiority of R-STDP over STDP. Second, we assessed the proposed network on several object recognition benchmarks.

\subsection{R-STDP Increases Computational Efficiency} 
Using STDP, when a neuron is exposed to input spike patterns, it tends to find the earliest repetitive sub-pattern by which the neuron reaches its threshold and fires~\cite{song2000competitive,guyonneau2005neurons,masquelier2008spike, masquelier2017stdp}. This tendency to favor early input spikes can be troublesome in case of distinguishing spike patterns that have temporal differences in their late sections.

\begin{figure}[!tb]
	\centering
	\par
	\subfloat[\label{fig:temporal_patterns}]{
		\includegraphics[width=5cm]{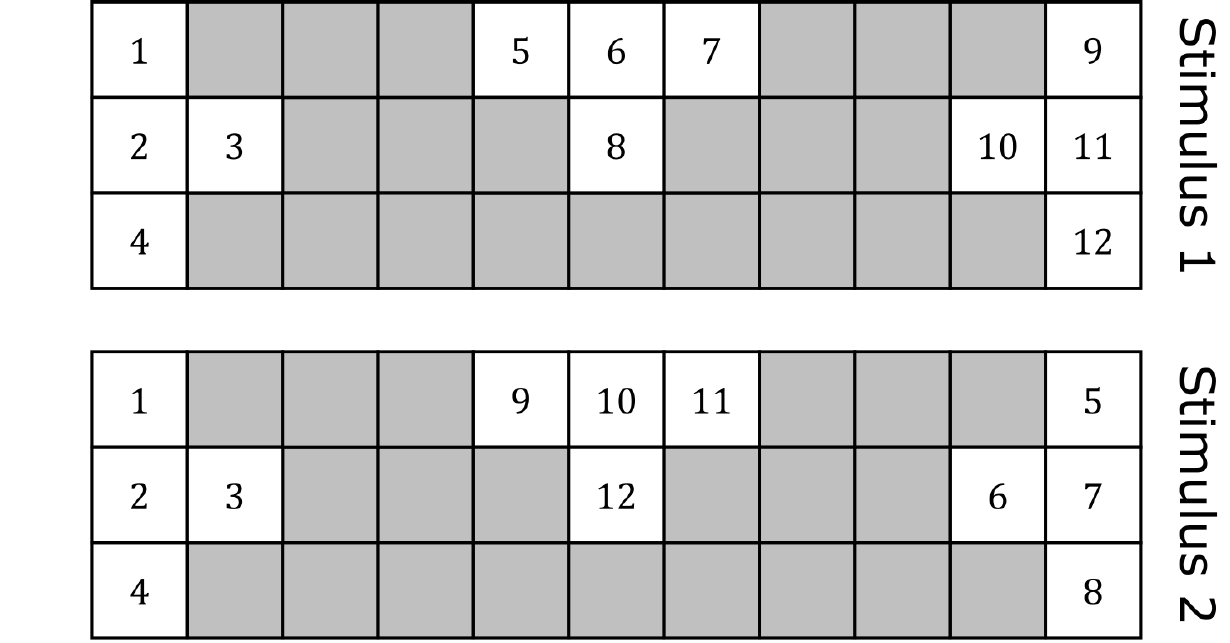}
	}\par
	\subfloat[\label{fig:temporal_features}]{
		\includegraphics[height=2cm]{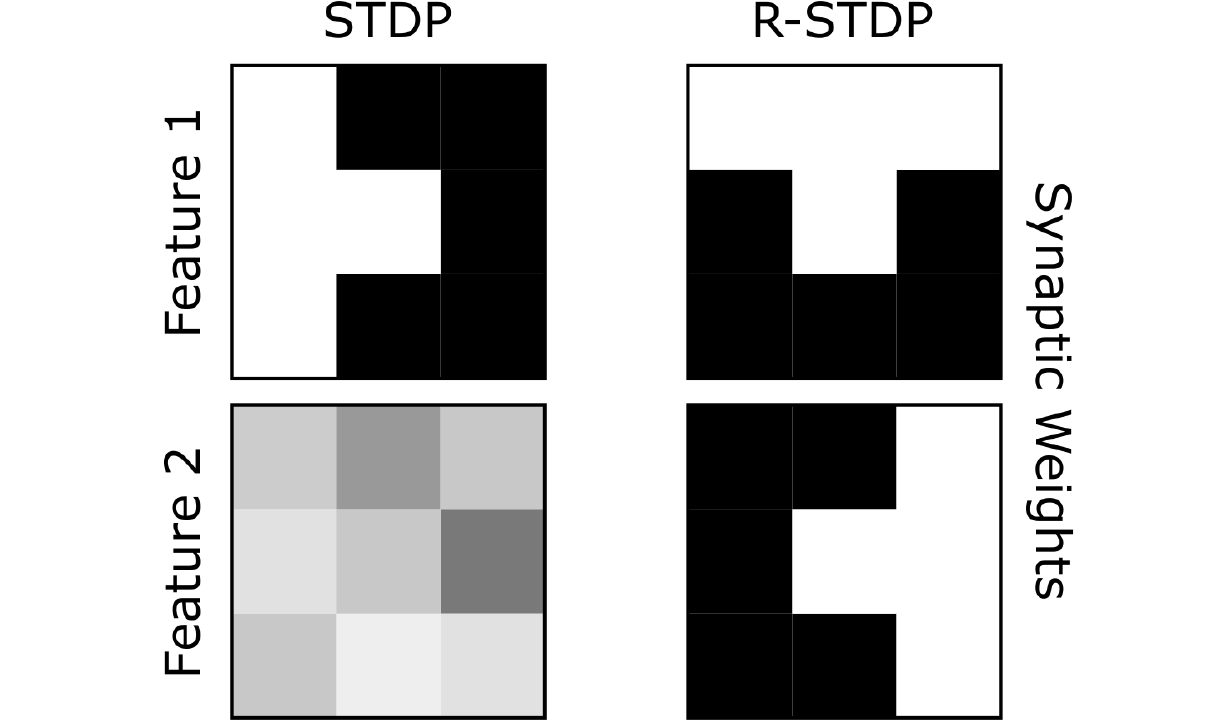}
	}
	\subfloat[\label{fig:temporal_features_big}]{
		\includegraphics[height = 2cm]{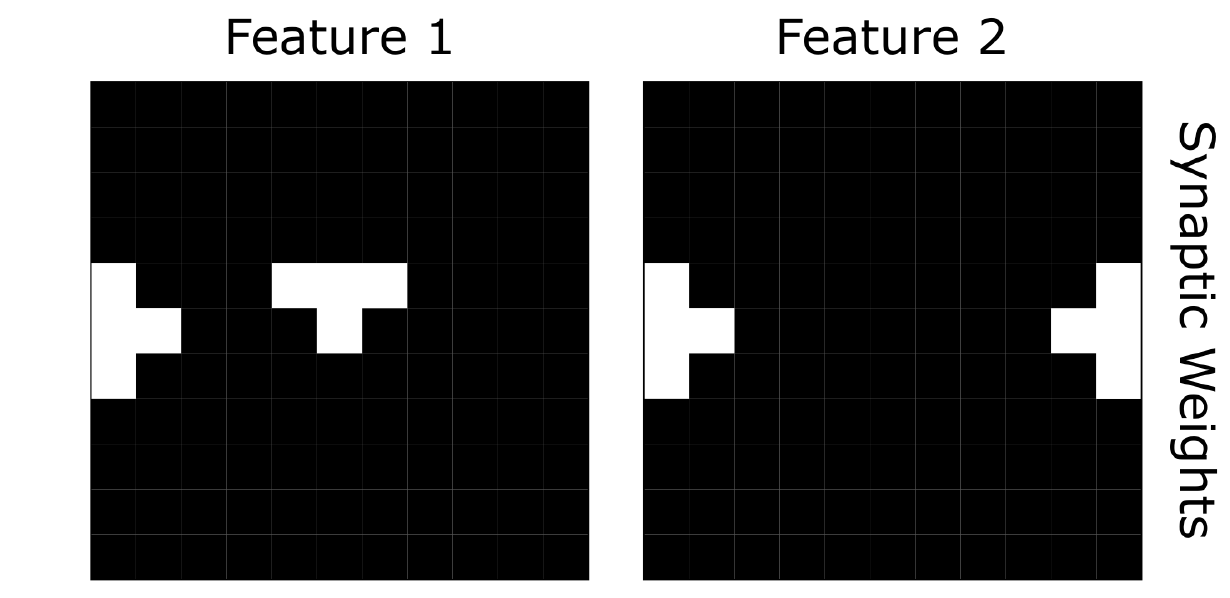}
	}	
	\caption{Temporal discrimination task. (a) Two input stimuli including a temporally different sub-pattern. Spikes are propagated from the white squares with the order written on them. (b) Synaptic weights (features) that are learned by the neurons with STDP (left column) and R-STDP (right column). The higher the weight the lighter is the gray level. (c) Synaptic weights when we used STDP-enabled neurons with large receptive fields and high thresholds.}\label{fig:temporal_disc}
\end{figure}

Assume that there are several categories of input stimuli that possess the same spatial configuration (Fig.~\ref{fig:temporal_patterns}). They also have identical early spikes. These patterns are repetitively presented to a group of IF neurons, for which the synaptic plasticity is governed by STDP and the one-winner-takes-all mechanism. If the neurons have low thresholds, one of them gets selective to the early common part of the input stimuli and inhibits the other neurons. Since the early parts are spatio-temporally the same among all of the input stimuli, there is no chance for the other neurons to fire and win the synaptic plasticity. Consequently, the overall activity of the neuronal group is the same for all of the input stimuli and classifies them into a single category.

As we will see below (Fig.~\ref{fig:temporal_features_big}), there are also some STDP-based solutions for this problem, however they are inefficient in using computational resources. For example, if we increase the size of receptive fields along with the thresholds, neurons gain the opportunity to receive the last spikes as well as the early ones. Another possible solution is to use many neurons that locally inhibit each other and drop the one-winner-takes-all constraint. This way, regarding the initial random weights, there is a chance for the neurons to learn other parts of the input stimuli.

Here, we show that the R-STDP learning rule solves this issue in a more efficient way than STDP. For this purpose, we designed an experiment containing two $3\times 11$ input stimuli. The inputs are spatially similar, which means that spikes are propagated from similar locations of both inputs. As illustrated in Fig.~\ref{fig:temporal_patterns}, each input is a 2D grid of white and gray squares. By white (gray) squares we denote locations, from which a spike is (is not) propagated. At the time of presenting any of these patterns to the network, spikes are propagated with a temporal order that is defined by the numbers written on the squares. According to this ordering, spikes with lower numbers are propagated earlier.

Since the input stimuli are artificial spike patterns, there was no need to apply Gabor filters, thus, they were fed directly into the layer $S2$. There, we put two neuronal grids with parameters $\omega_{s2} = 3$ and $\mathcal{T} = 3$. Therefore, each grid contained $1\times 9$ neurons to cover the entire input stimuli. We also set $a_r^+ = 0.05, a_r^- = -0.05, a_p^+ = 0.1$, and $a_p^- = -0.1$. The goal of the task was that the first (second) $C2$ neuron fires earlier for the first (second) pattern. We examined both STDP and R-STDP learning rules to see if the network finds discriminative features or not.

As shown in Fig.~\ref{fig:temporal_features}, using STDP, the network extracted a non-discriminative feature, the shared one between both input stimuli. On the other side, the proposed reinforcement learning mechanism guided the neurons to extract features whose temporal order of appearance is the only thing leading to a successful pattern discrimination. We repeated this experiment for $100$ times using different random initial weights. Results showed that our network succeeded in $98\%$ of the times, while there were no chance for STDP to find the discriminative features. When we increased the threshold to $4$ (requiring at least two sub-patterns) and the size of the receptive fields to $11\times 11$ (covering the entire pattern), the network employing the STDP could also find discriminative features (see Fig.~\ref{fig:temporal_features_big}) in 80\% of the times.

\subsection{Plastic Neurons}
As mentioned earlier, the brain reward system plays an important role in the emergence of a particular behavior. In this part of the paper, we demonstrate the R-STDP's capability of re-adjusting neurons' behavior in an online manner.

\begin{figure*}[tb]
	\centering
	\par
	\subfloat[\label{fig:behavior_a}]{
		\includegraphics[width=12cm]{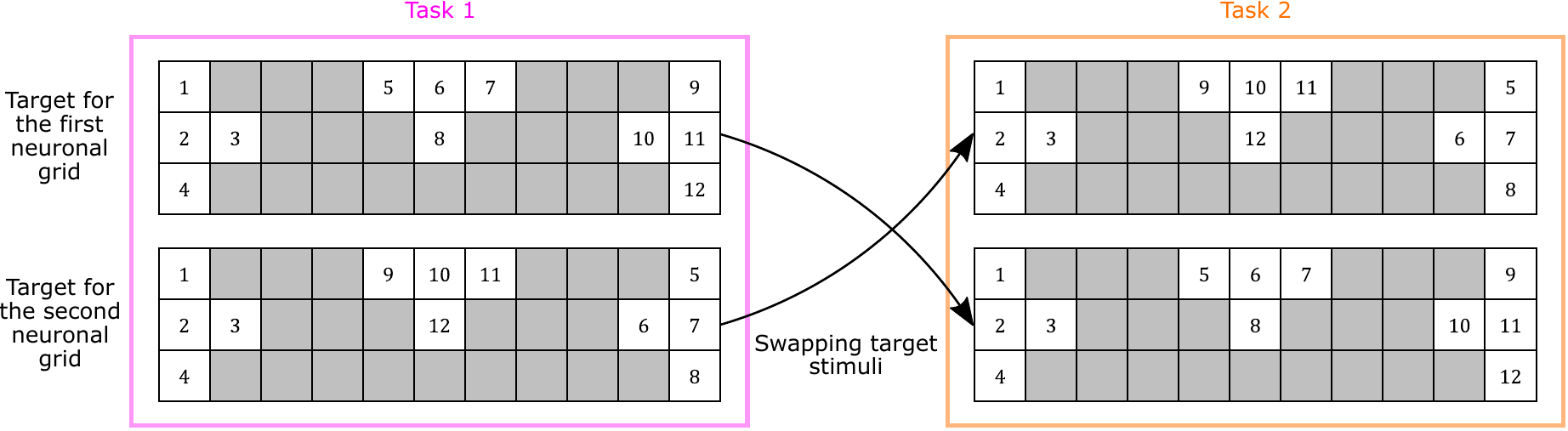}
	}\par
	\subfloat[\label{fig:behavior_b}]{
		\includegraphics[width=15cm]{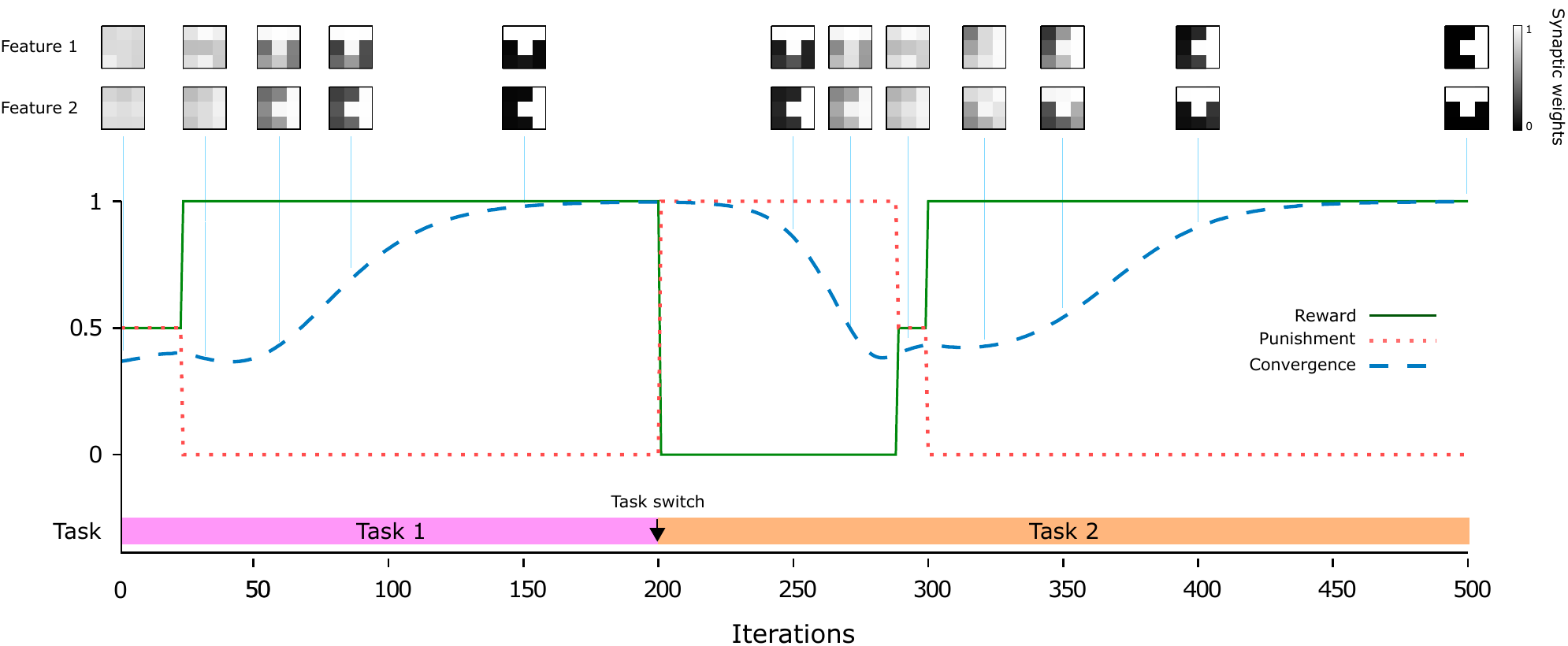}
	}
	\caption{Neurons with flexible behavior. (a) Target stimuli for each neuronal grid. In Task 2, the target stimuli of Task 1 are swapped. (b) Flexibility of the network in a changing environment. Plots on the top two rows represent changes of the synaptic weights. The plot on the bottom illustrates changes in the rate of receiving reward (green solid line), punishment (red dotted line), and the convergence of the synaptic weights (blue dashed line) over $500$ iterations. Convergence is measured by $1 - \frac{\sum_{}^{}W_{ij}\left(1-W_{ij}\right)}{18}$, where the summation is over the $18$ ($3\times 3 + 3\times 3$) synaptic weights that are shared between $S2$ neurons. As this value gets closer to one, the synaptic weights have more binary-like values.}\label{fig:behavior}
\end{figure*}

We designed an experiment, in which the pre-defined desired behavior of the neurons is changed during the simulation. The experimental setup is very similar to the ``Temporal Discrimination" task with similar input stimuli and parameter values, except that we swapped the target input stimuli during the training iterations (see Task 1 and Task 2 in Fig.~\ref{fig:behavior_a}). As shown in Fig.~\ref{fig:behavior_b}, at the beginning of the simulation, the desired behavior was that the neurons belonging to the first grid respond to the first stimulus earlier than those in the second grid, and vice versa. After $200$ iterations, when the convergence is fulfilled, we swapped the target stimuli. At this stage, since the neurons were exclusively sensitive to the previous target stimuli, they began to generate false alarms. Consequently, the network was receiving high rates of punishments for around $80$ iterations (see iterations $200$ to $280$ in Fig~\ref{fig:behavior_b}), which in turn swapped LTD and LTP (see Materials and Methods: Reward-modulated STDP). As the network received punishments, the previously weakened (strengthened) synapses got stronger (weaker). Therefore, the sensitivity diminished for a while, and the neurons regained the possibility of learning something new. After iteration $300$, neurons found their new target stimulus and, once again, converged to the discriminative features (see the plots of synaptic weights in the top two rows in Fig.~\ref{fig:behavior_b}).

In summary, R-STDP enables the neurons to unlearn what they have learned so far. This ability results in neurons with flexible behavior (plastic neurons), that are able to learn rewarding behavior in changing environments. This ability also helps the neurons to forget and escape from the local optima in order to learn something that earns more reward. Applying STDP in such a scenario does not work at all, since there is no difference between Task $1$ and Task $2$ from an unsupervised point of view.

\subsection{Object Recognition}
In this section, the performance of our network on categorization of natural images is evaluated. We begin with a description of the datasets that are used in our experiments. Then, we show how the network benefits from the reinforcement learning mechanism to extract features from natural images, followed by comparing R-STDP and STDP in object recognition tasks. Finally, we illustrate how the dropout and adaptive learning techniques reduce the chance of overfitting to the training samples.

\subsubsection{Datasets}
We used three well-known object recognition benchmarks to evaluate the performance of the proposed network. The first and easiest one is Caltech face/motorbike which is mainly used for demonstration purposes. The next two that are used to evaluate the proposed network are ETH-80 and small NORB. These datasets contain images of objects from different view points which make the task harder (see supplementary Fig. S1). 

\subsubsection{Reinforced Selectivity}
The previous experiments showed that R-STDP enables the network to find informative and discriminative features, both spatially and temporally. Here, we show that R-STDP encourages the neurons to become selective to a particular category of natural images. To this end, we trained and examined the network on images from two categories of face and motorbike from the Caltech dataset.

In this experiment, we put $10$ neuronal grids for each category that were reinforced to win the first-spike competition in response to the images from their target categories. Therefore, the desired behavior of the network was that the neurons of the first $10$ grids get selective to the face category, while those in the other grids get selective to the motorbikes.

\begin{figure*}[htb!]
	\centering
	\par
	\subfloat[\label{fig:caltech_demo_a}]{
		\includegraphics[width = 15cm]{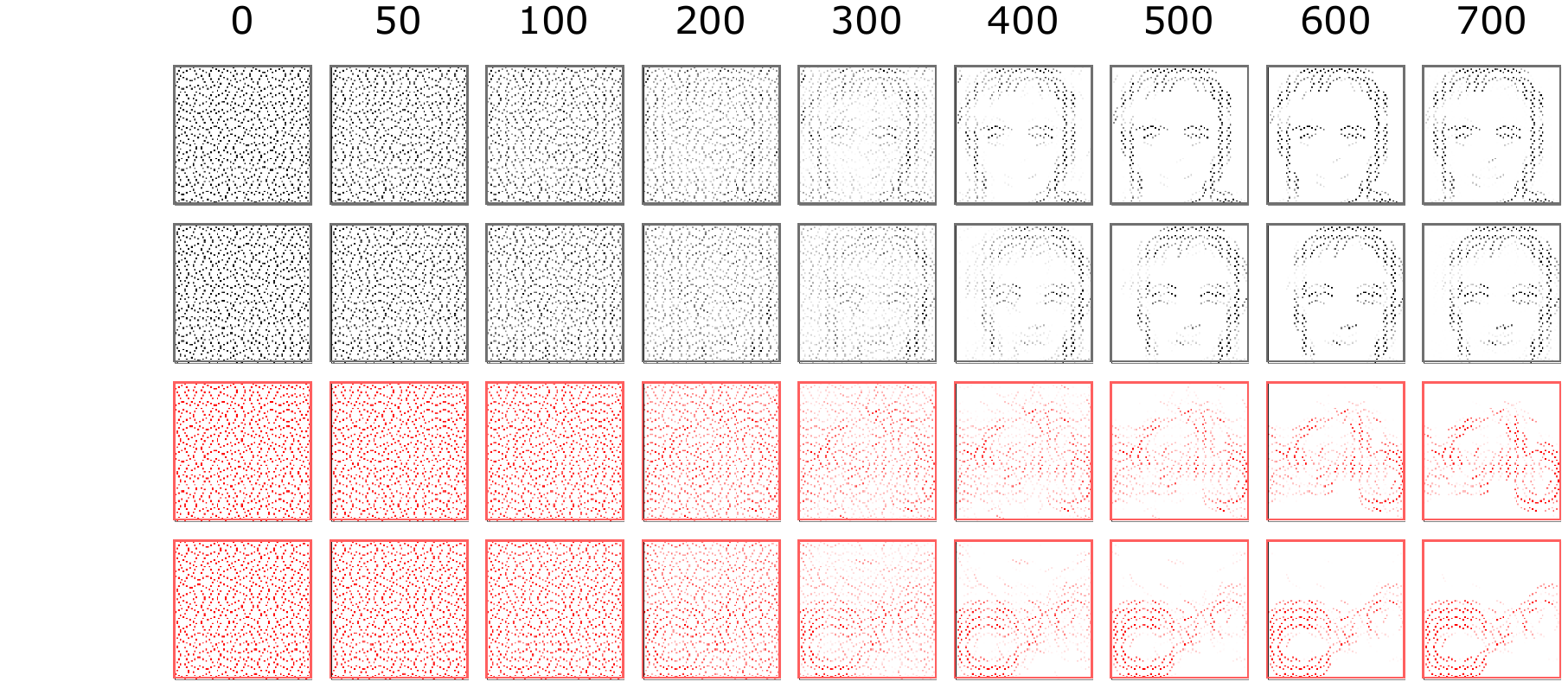}
	}
	\par
	\subfloat[\label{fig:caltech_demo_b}]{
		\includegraphics[width = 15cm]{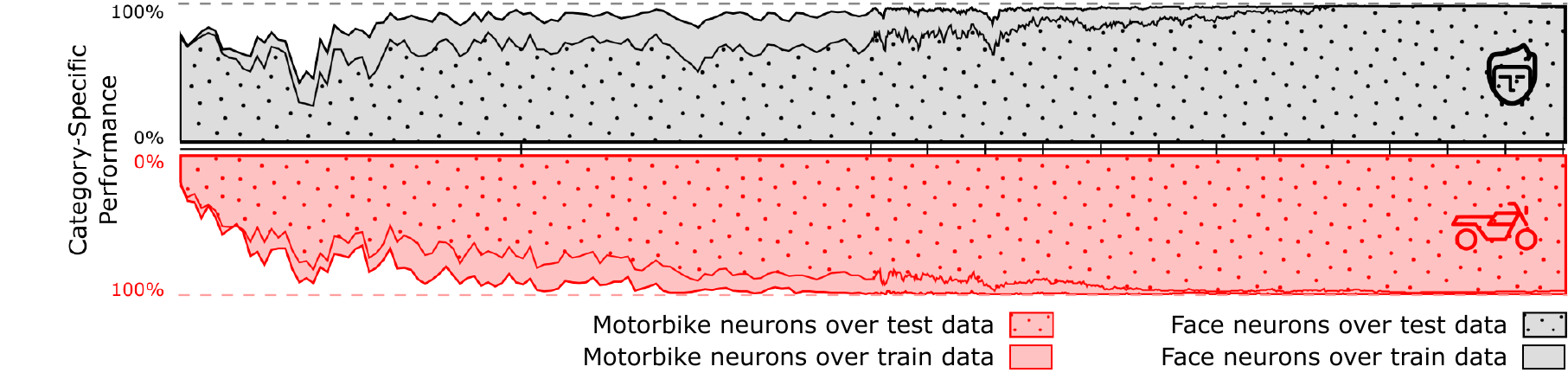}
	}
	\par
	\subfloat[\label{fig:caltech_demo_c}]{
		\includegraphics[width = 15cm]{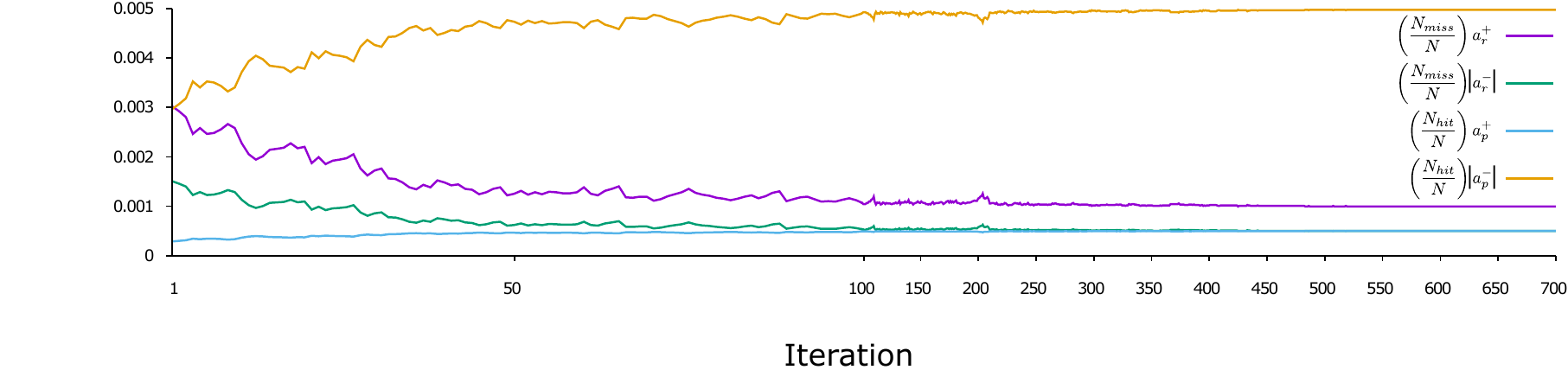}
	}
	\caption{Training the network on Caltech face/motorbike dataset. (a) Evolution of four different features (out of $20$) extracted by the network. The black and red plots correspond to the face and motorbike neurons, respectively. (b) Hit rate for neurons of each category. The gray (pink) filled curves depict the percentage of the times that the face (motorbike) neurons emit the earliest spike in response to their target stimulus. Notice that curves for motorbike neurons are mirrored vertically for the sake of better illustration, and hit rates over testing set are indicated by dot-patterns. (c) Trajectory of changes in learning rate with respect to the number of correct ($N_{hit}$) and incorrect ($N_{miss}$) categorizations.}
	\label{fig:caltech_demo}
\end{figure*}

Fig.~\ref{fig:caltech_demo} illustrates the behavior of the network over the training iterations. Since the early iterations contained rapid changes, they are plotted wider. During early iterations, strong synaptic weights (see Materials and Methods: Layer $S2$) and $50\%$ dropout probability resulted in an unstable network  whose neurons responded to random input stimuli. This chaotic behavior can be easily spotted on early iterations in the middle plot (see Fig.~\ref{fig:caltech_demo_b}). As the network continues training iterations, reward/punishment signals made neurons more and more selective to their target categories. As shown in Fig.~\ref{fig:caltech_demo_b}, after $200$ iterations, a quite robust selectivity appeared for the training samples, while on the testing samples, it is elongated for $300$ more iterations. This quick convergence on training samples is due to the fact that the network is relatively fast in finding features that successfully discriminate seen samples (see Fig.~\ref{fig:caltech_demo_a}). These primary features need to converge more to be applicable on testing samples, which requires even more iterations because of the adaptive learning rates. Moreover, we do not let the learning rate drops below $20\%$ of the values of parameters $a_r^+$, $a_r^-$, $a_p^+$, and $a_p^-$. This allows the network to continue convergence with a constant rate even if all of the training samples are correctly categorized (see Fig.~\ref{fig:caltech_demo_c}).

We repeated the experiment $30$ times with random initial weights and different training and testing samples and the performance achieved by the proposed network is $98.9\pm 0.4\%$ (mean $\pm$ std). When we tried a same network structure with STDP, $97.2\%$ was its best achievement (see Table~\ref{tbl:perf_comp}).

\subsubsection{Performance}
We have shown how the proposed network successfully classified faces from motorbikes with high accuracy. Here, we examined the performance of the proposed network on the ETH-80 and NORB datasets that are more challenging (see Supplementary Materials: Datasets). The performance of the network is tested over the entire testing set after each training iteration, in which the network receives all of the training samples in random order.

For ETH-80 dataset, we configured the network to extract $10$ features per category, which resulted in $8 \times 10 = 80$ features in total. The receptive field of each neuron in layer $S2$ was set in a way that it covered the whole input image. Here, nine instances of each category were presented to the network as the training samples, and the remaining were employed in the test phase. After performing $250$ training and testing iterations, the best testing performance of the network was reported.

Again, we repeated this experiment $30$ times, each time using a different training and testing set. As before, the network successfully extracted discriminative features (see Supplementary Fig. 2) and reached the performance of $89.5\pm 1.9\%$ (mean $\pm$ std). We also applied STDP to a network with the same structure. To examine the STDP performance, we used support vector machines with linear kernel and KNNs (K was changed from $1$ to $10$). According to the results, the accuracy achieved by this network is $84.5\%$, when the maximum potentials were used as the feature vectors and the classifier was KNN. Considering that the proposed network classifies input patterns solely based on the first-spike information, R-STDP definitely outperforms STDP. Table~\ref{tbl:perf_comp} provides the details of the comparison made between R-STDP and STDP.

By looking at confusion matrices (see Supplementary Fig. 3a), we found that both R-STDP and STDP agree on the most confusing categories, that are cow, dog, and horse. However, thanks to the reinforcement learning, R-STDP not only decreased the confusion error, but also provided a more balanced error distribution.

\begin{table*}
	\begin{center}
		\caption{Comparison of the network's performance when using R-STDP and STDP.}
		\label{tbl:perf_comp}
		\begin{tabular}{|l|r|r|r|r|r|r|r|r|}
			\hline
			\multicolumn{1}{|c|}{\multirow{3}{*}{Dataset}}
			&
			\multicolumn{1}{c|}{\multirow{3}{*}{R-STDP}} &
			\multicolumn{6}{c|}{STDP} & 
			\multicolumn{1}{c|}{\multirow{3}{*}{Shallow CNN}}\\
			\cline{3-8}
			&&\multicolumn{2}{c|}{First-Spike}&\multicolumn{2}{c|}{Spike-Count}&\multicolumn{2}{c|}{Max-Potential}&\\
			\cline{3-8}
			&& \multicolumn{1}{c|}{SVM} & \multicolumn{1}{c|}{KNN} & \multicolumn{1}{c|}{SVM} & \multicolumn{1}{c|}{KNN} & \multicolumn{1}{c|}{SVM} & \multicolumn{1}{c|}{KNN} &\\
			\hline
			Caltech (Face/Motorbike) & $98.9$ & $96.4$ & $96.4$ & $96.9$ & $93.4$ & $96.6$ & $97.2$ & $99.3$ \\
			ETH-80 & $89.5$ & $72.9$ & $69.8$ & $74$ & $70.4$ & $79.9$ & $84.5$ & $87.1$\\
			NORB & $88.4$ & $62.7$ & $58.6$ & $61.7$ & $55.3$ & $66$ & $65.9$ & $85.5$ \\
			\hline
		\end{tabular}
	\end{center}
\end{table*}

The same experiment was also performed on the NORB dataset. Again, we put $10$ neuronal grids for each of the five categories, whose neurons are able to see the entire incoming stimuli. The proposed network with R-STDP reached the performance of  $88.4\pm 0.5\%$ (mean $\pm$ std) on testing samples, whereas STDP achieved $66\%$ at most. By reviewing confusion matrices of both methods, we found that both networks encountered difficulties mostly in distinguishing four-leg animals from humans, as well as cars from trucks (see Supplementary Fig. 3b). As before, R-STDP resulted in a more balanced error distribution.

Additionally, we compared the proposed network to convolutional neural networks (CNNs). Although the proposed network is not able to beat pre-trained deep CNNs such as VGG16~\cite{simonyan2014very} (see Supplementary Materials: Comparison with Deep Convolutional Neural Networks), comparing it to a shallow CNN, with a similar network structure and same input would be a fair point. We repeated all of the object categorization experiments using a shallow CNN implemented with Keras neural networks API and Tensorflow as its backend. As shown in Table~\ref{tbl:perf_comp}, the proposed network successfully outperformed the supervised CNN in both of the ETH-80 and NORB datasets.

\subsubsection{Overfitting Problem}
Overfitting is one of the most common issues in supervised or reinforcement learning scenarios. This problem got even worse by the emergence of deep learning algorithms. There are many studies focused on developing techniques that increase the generalization power of the learning algorithms. One of the mechanism that has shown promising empirical results on deep neural networks is the dropout technique~\cite{krizhevsky2012imagenet}. This technique temporary reduces the complexity of the network by suppressing the activity of a specific number of neurons. This reduction in neuronal resources forces the network to generalize more in order to reduce the prediction error.

\begin{figure}[h]
	\begin{center}
		\includegraphics[width = 8.7cm]{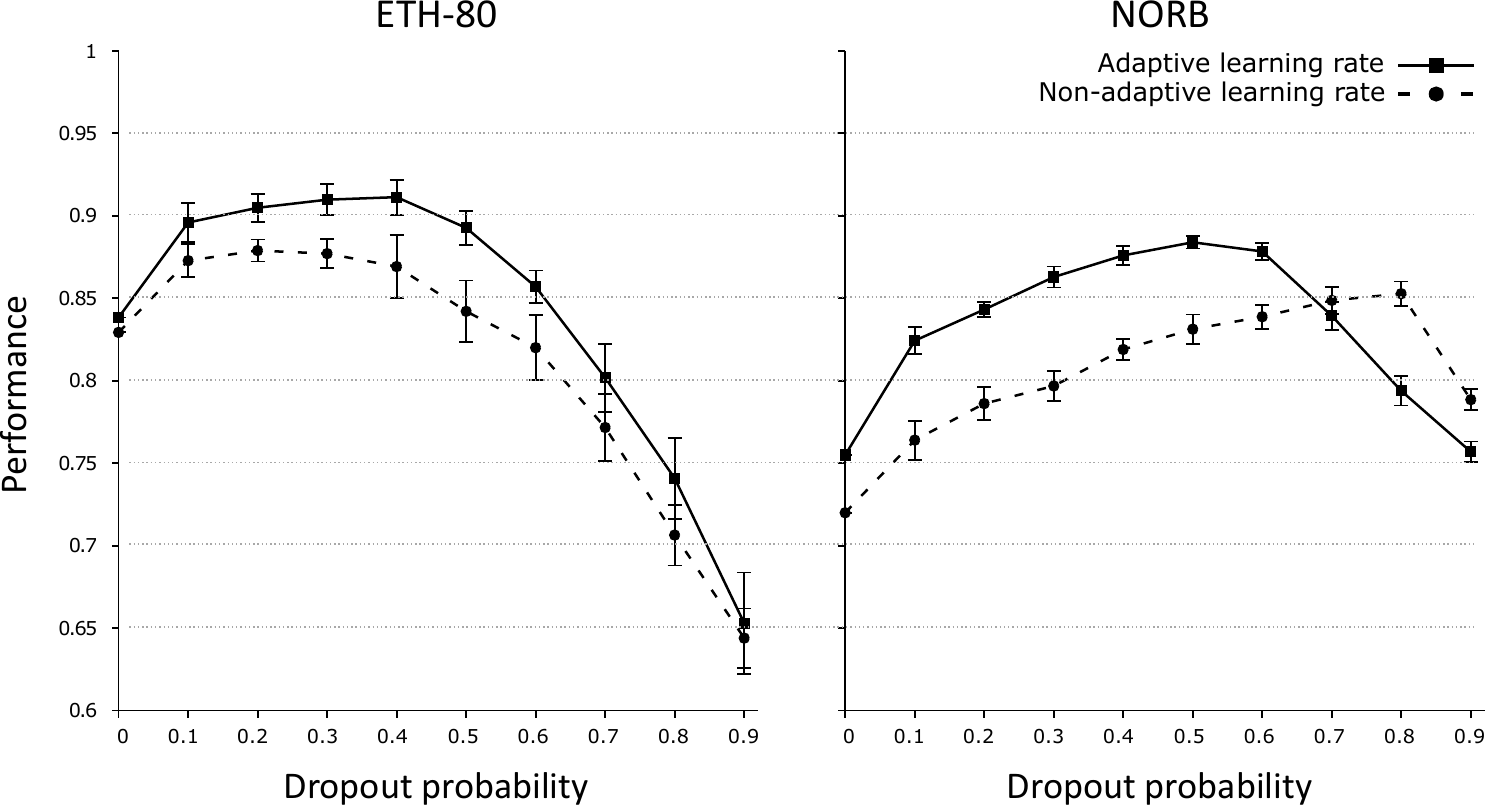}
		\caption{Impact of the dropout and the adaptive learning rate techniques. The plot on the left (right) demonstrates the result for ETH-80 (NORB) dataset. In these plots the solid (dashed) lines illustrate the performance of the network with different dropout probabilities when the adaptive learning rate is on (off).}
		\label{fig:overfit}
	\end{center}
\end{figure}

The proposed network is not an exception and has shown tendencies to overfit on the training samples through our examinations. Therefore, we adopted the dropout technique in our experiments. We also found that an steady learning rate does increase the chance of overfitting. Thus, we made use of dynamic learning rates with respect to the performance of the network (see Material and Methods: Overfitting Avoidance).

To show the impact of the aforementioned mechanisms, we repeated the object recognition experiments with different dropout probabilities and steady learning rates. Fig.~\ref{fig:overfit} simultaneously shows the impact of both mentioned mechanisms on categorization of test samples. It is clear that when the adaptive learning rate mechanism is applied, the network achieved higher performances (solid lines). It is also shown that the dropout probability must be chosen according to the complexity of the dataset as well as the network. Since the NORB dataset contains more complex samples than the ETH-80, it tends more to overfitting on training samples. As a consequence, it needs more dropout rate to overcome this issue. The magnitude of this tendency is even clearer when the steady learning rates are used. To put it differently, faster convergence rate along with the complexity of the samples induce more overfitting, which in turn needs more dropout rate.

\section{Discussion}
Mammals are fast and accurate at visual object recognition. Their visual cortex processes the incoming data in a hierarchical manner, through which the complexity of neuronal preference is gradually increased. This hierarchical processing provides a robust and invariant object recognition~\cite{thorpe1996speed,hung2005fast,dicarlo2007untangling,liu2009timing, dicarlo2012does}. Computational modeling of the mammalian visual cortex has been under investigation for many years. Developing a biologically plausible model not only enables scientists to examine their hypotheses with low cost, but also provides a human-like vision for artificially intelligent machines~\cite{masquelier2007unsupervised,fukushima1982neocognitron,lecun1995convolutional,serre2007robust,lee2009convolutional}.

Deep convolutional neural networks (DCNNs) are the most successful works in this area~\cite{krizhevsky2012imagenet,zeiler2014visualizing,simonyan2014very, kheradpisheh2016humans, kheradpisheh2016deep}. The idea behind these networks is inspired by the hierarchical structure of the visual cortex. Despite the promising results obtain by DCNNs, they are not biologically plausible because of using supervised learning rules. In addition, they employ rate-based encoding scheme, which is both energy and resource consuming. There is another group of studies trying to use spiking neurons along with the unsupervised STDP learning rule~\cite{masquelier2007unsupervised, querlioz2013immunity,kheradpisheh2017stdp, diehl2015unsupervised, tavanaei2016acquisition}. These models are more biologically plausible, but they cannot beat DCNNs in terms of accuracy. In theory, spiking neural networks (SNNs) have more computational power than DCNNs, however, they are harder to control because of the complex dynamics and high dimensional space of effective parameter. Furthermore, since most of them are trained in an unsupervised manner, the classification step is done by an external classifier or statistical methods.

Here, we solved the object recognition task using a hierarchical SNN equipped with a reinforcement learning rule called R-STDP~\cite{fremaux2016neuromodulated}. There are several studies showing that the brain uses RL to solve the problem of decision-making~\cite{niv2009reinforcement, lee2012neural,steinberg2013causal, schultz2015neuronal}. Therefore, it is a suitable choice for training class-specific neurons that are able to decide on the class of the input image. Therefore, we put one step further developing a more biologically plausible model which is able to perform the visual categorization totally on its own. The proposed network functions in the temporal domain, where the information is encoded by spike times. The input image is first convolved with oriented Gabor filters and a spike train is generated based on a latency-to-intensity coding scheme. The resulting spikes are then propagated toward the feature extraction layer. Using R-STDP, the proposed network successfully found task-specific diagnostic features using neurons that were pre-assigned to the class labels. In other words, each neuron was assigned to a class a priori, where its desired behavior was to respond early for the instances belonging to the specified class. To decrease the computational cost even more, neurons were forced to fire at most once for an input image and the latency of their spike is considered as the measure of stimulus preference. Therefore, if a neuron fired earlier than the others, it would have received its preferred stimulus. This measure of preference served as an indicator for the network's decision. That is to say, when a neuron belonging to a particular class fired earlier, the network's decision was considered to be that class.

Through our experiments, we compared R-STDP to STDP from different aspects. We showed that R-STDP can save computational resources. This was clarified by a hand-designed discrimination task, in which the order of spikes was the only discriminative feature. R-STDP solved the problem using minimal number of neurons, synapses, and threshold, whereas STDP needed more neurons, more synapses, and higher thresholds. This drawback for STDP is due to the fact that it tends to find statistically frequent features~\cite{masquelier2008spike, gilson2011stdp, brette2012computing, masquelier2017stdp}, which are not necessarily the diagnostic ones. As a consequence, one needs to use either more neurons or more synapses to ensure that the diagnostic features will be eventually found. On the other hand, since R-STDP informs the neurons about their outcomes, they can function better using minimal resources.

After having demonstrated the advantages of R-STDP in finding diagnostic features, we investigated how well it can be combined with a hierarchical SNN for solving both visual feature extraction and object categorization in a biologically plausible manner. We evaluated the proposed network and a similar network which uses STDP, as well as a CNN with the same structure, on three datasets of natural images Caltech Face/Motorbike, ETH-80 and NORB. The last two contain images of objects from different viewpoints, which made the task harder. When we compared the performances obtained by the networks, we found that R-STDP strongly outperforms STDP and the CNN with same structure. An even more interesting point is that the proposed network achieved this superiority decisions solely based on the first-spikes, while in the case of the others, even the powerful classifiers like SVMs and error back-propagation were not of any help.

To compare R-STDP with STDP, both networks used the same values for parameters except the learning rate (see Materials and Methods: Comparison of the R-STDP and STDP). However, one can use STDP with higher number of neurons and tuned thresholds to compensate the blind unsupervised feature extraction and achieve better performances~\cite{kheradpisheh2016bio}. Again, we conclude that R-STDP helps the network to act more efficiently in consuming computational resources.

Putting everything together, the proposed network has the following prominent features:
\begin{itemize}
	\item Robust object recognition in natural images.
	\item Each neuron is allowed to spike only once per image. This results in a huge reduction of energy consumption.
	\item Decision-making (classification) is performed using the first-spike latencies instead of powerful classifiers. Therefore, the biological plausibility of the model is increased.
	\item Synaptic plasticity is governed by RL (the R-STDP rule), for which supporting biological evidence can be found~\cite{fremaux2016neuromodulated}, and which allows to extract highly diagnostic features.
\end{itemize}

Our network can be interesting for neuromorphic engineering~\cite{furber2016large}, since it is both biologically plausible and hardware-friendly. Although hardware implementation and efficiency is out of the scope of the current paper, we believe that the proposed network can be implemented in hardware in an energy-efficient manner for several reasons. Firstly, SNNs are more hardware friendly than classic artificial neural networks, because the energy-consuming ``multiply-accumulator" units can be replaced by more energy-efficient ``accumulator" units. For this reason, studies on training deep convolutional SNNs (DCSNNs)~\cite{lee2016training,kheradpisheh2017stdp} and converting DCNNs into DCSNNs~\cite{ruckauer2017conversion}, as well as restricted DCNNs~\cite{courbariaux2015binaryconnect, binas2016deep,esser2016convolutional} have gained interests in recent years. Secondly, most SNN hardwares use event-driven approaches by considering spikes as events. This way, energy consumption increases with the number of spikes. Thus, by allowing at most one spike per neuron, the proposed model is as efficient as possible.  Finally, the proposed learning rule is more suitable for online, on-chip learning than error backpropagation in deep networks, where updating weights based on high-precision gradients brings difficulties for hardware implementation.

To date, we could not find any other works possessing the aforementioned features. To mention one of the closest attempts, Gardner et al.~\cite{gardner2014classifying} tried to classify Poisson-distributed spike trains by a readout neuron equipped with R-STDP. Although their method is working, it cannot be applied on natural images as it is, because of their time-based encoding and target labeling. There is another related work by Huerta and Nowotny~\cite{huerta2009fast}. In this work, the authors designed a model of the RL mechanism which occurs in the mushroom body. They applied their RL mechanism on a pool of randomly connected neurons with $10$ readout neurons to classify handwritten digits. Our work is different from theirs in several aspects. First, we used a hierarchical structure based on the mammalian visual cortex, while they used randomly connected neurons. Second, we used the R-STDP learning rule, whereas they employed a probabilistic approach for the synaptic plasticity. Third, the input of our network were natural images using intensity-to-latency encoding, while they used binary encoding with a threshold on artificial images.

Although the results of the proposed network were significantly better than the network employing STDP with external classifiers, they are still not competitive to the state-of-the-art deep learning approaches. One of the limitations to the current method is using only one trainable layer. Besides, the receptive field of the neurons in the last layer are set to be large enough to cover an informative portion of the image. As a result, the network cannot resist high rates of variations in the object, unless using more and more number of neurons. Extending the number of layers in the current network is one of the directions for future research. Going deeper seems to improve the performance by providing a gradual simple to complex feature extraction. However, deeper structure needs more parameter tuning, and a suitable multi-layer synaptic plasticity rule. Recent studies have also shown that combining deep networks and RL can lead to outstanding results~\cite{mnih2015human, silver2016mastering}.

Another direction for the future research is to use the RL for learning semantic associations. For example, STDP is able to extract features for different kinds of animals in different viewpoints, but it is not able of relating all of them into the category of ``animal", because different animals have no reason to co-occur. Or, it can extract features for the frontal and profile face, but it cannot generate an association putting both in the general category of ``face". On the other hand, by a reinforcement signal and using learning rules like R-STDP, neurons are not only able to extract diagnostic features, but also learn relative connections between categories and create super-categories.

\section*{Acknowledgment}
The authors would like to thank Dr. Jean-Pierre Jaffr\'ezou for proofreading this manuscript, as well as Amirreza Yousefzadeh and Dr. Bernab\'e Linares-Barranco for providing useful hardware-related information.

This research received funding from the European Research Council under the European Union's 7th Framework Program (FP/2007–2013)/ERC Grant Agreement no. 323711 (M4 project), Iran National Science Foundation: INSF (No. 96005286), and Institute for Research in Fundamental Sciences (BS-1396-02-02), Tehran, Iran.

\footnotesize

\begin{thebibliography}{10}
	\expandafter\ifx\csname url\endcsname\relax
	\def\url#1{\texttt{#1}}\fi
	\expandafter\ifx\csname urlprefix\endcsname\relax\def\urlprefix{URL }\fi
	\expandafter\ifx\csname href\endcsname\relax
	\def\href#1#2{#2} \def\path#1{#1}\fi
	
	\bibitem{gerstner1996neuronal}
	W.~Gerstner, R.~Kempter, J.~L. van Hemmen, H.~Wagner, A neuronal learning rule
	for sub-millisecond temporal coding, Nature 383~(6595) (1996) 76.
	
	\bibitem{markram1997regulation}
	H.~Markram, J.~L{\"u}bke, M.~Frotscher, B.~Sakmann, Regulation of synaptic
	efficacy by coincidence of postsynaptic {AP}s and {EPSP}s, Science 275~(5297)
	(1997) 213--215.
	
	\bibitem{bi1998synaptic}
	G.-q. Bi, M.-m. Poo, Synaptic modifications in cultured hippocampal neurons:
	dependence on spike timing, synaptic strength, and postsynaptic cell type,
	Journal of Neuroscience 18~(24) (1998) 10464--10472.
	
	\bibitem{sjostrom2001rate}
	P.~J. Sj{\"o}str{\"o}m, G.~G. Turrigiano, S.~B. Nelson, Rate, timing, and
	cooperativity jointly determine cortical synaptic plasticity, Neuron 32~(6)
	(2001) 1149--1164.
	
	\bibitem{meliza2006receptive}
	C.~D. Meliza, Y.~Dan, Receptive-field modification in rat visual cortex induced
	by paired visual stimulation and single-cell spiking, Neuron 49~(2) (2006)
	183--189.
	
	\bibitem{huang2014associative}
	S.~Huang, C.~Rozas, M.~Trevi{\~n}o, J.~Contreras, S.~Yang, L.~Song,
	T.~Yoshioka, H.-K. Lee, A.~Kirkwood, Associative {Hebbian} synaptic
	plasticity in primate visual cortex, Journal of Neuroscience 34~(22) (2014)
	7575--7579.
	
	\bibitem{guo2017timing}
	Y.~Guo, W.~Zhang, X.~Chen, J.~Fu, W.~Cheng, D.~Song, X.~Qu, Z.~Yang, K.~Zhao,
	Timing-dependent {LTP} and {LTD} in mouse primary visual cortex following
	different visual deprivation models, PLoS One 12~(5) (2017) e0176603.
	
	\bibitem{masquelier2008spike}
	T.~Masquelier, R.~Guyonneau, S.~J. Thorpe, Spike timing dependent plasticity
	finds the start of repeating patterns in continuous spike trains, {PLoS} one
	3~(1) (2008) e1377.
	
	\bibitem{gilson2011stdp}
	M.~Gilson, T.~Masquelier, E.~Hugues, {STDP} allows fast rate-modulated coding
	with poisson-like spike trains, {PLoS} Computational Biology 7~(10) (2011)
	e1002231.
	
	\bibitem{brette2012computing}
	R.~Brette, Computing with neural synchrony, {PLoS} Computational Biology 8~(6)
	(2012) e1002561.
	
	\bibitem{masquelier2017stdp}
	T.~Masquelier, Stdp allows close-to-optimal spatiotemporal spike pattern
	detection by single coincidence detector neurons, Neuroscience.
	
	\bibitem{sutton1998introduction}
	R.~S. Sutton, A.~G. Barto, Introduction to reinforcement learning, Vol. 135,
	MIT Press Cambridge, 1998.
	
	\bibitem{dayan2002reward}
	P.~Dayan, B.~W. Balleine, Reward, motivation, and reinforcement learning,
	Neuron 36~(2) (2002) 285--298.
	
	\bibitem{daw2006computational}
	N.~D. Daw, K.~Doya, The computational neurobiology of learning and reward,
	Current Opinion in Neurobiology 16~(2) (2006) 199--204.
	
	\bibitem{niv2009reinforcement}
	Y.~Niv, Reinforcement learning in the brain, Journal of Mathematical Psychology
	53~(3) (2009) 139--154.
	
	\bibitem{lee2012neural}
	D.~Lee, H.~Seo, M.~W. Jung, Neural basis of reinforcement learning and decision
	making, Annual Review of Neuroscience 35 (2012) 287--308.
	
	\bibitem{steinberg2013causal}
	E.~E. Steinberg, R.~Keiflin, J.~R. Boivin, I.~B. Witten, K.~Deisseroth, P.~H.
	Janak, A causal link between prediction errors, dopamine neurons and
	learning, Nature Neuroscience 16~(7) (2013) 966--973.
	
	\bibitem{schultz2015neuronal}
	W.~Schultz, Neuronal reward and decision signals: from theories to data,
	Physiological Reviews 95~(3) (2015) 853--951.
	
	\bibitem{schultz2002getting}
	W.~Schultz, Getting formal with dopamine and reward, Neuron 36~(2) (2002)
	241--263.
	
	\bibitem{schultz1998predictive}
	W.~Schultz, Predictive reward signal of dopamine neurons, Journal of
	Neurophysiology 80~(1) (1998) 1--27.
	
	\bibitem{glimcher2011understanding}
	P.~W. Glimcher, Understanding dopamine and reinforcement learning: the dopamine
	reward prediction error hypothesis, Proceedings of the National Academy of
	Sciences 108~(Supplement 3) (2011) 15647--15654.
	
	\bibitem{seol2007neuromodulators}
	G.~H. Seol, J.~Ziburkus, S.~Huang, L.~Song, I.~T. Kim, K.~Takamiya, R.~L.
	Huganir, H.-K. Lee, A.~Kirkwood, Neuromodulators control the polarity of
	spike-timing-dependent synaptic plasticity, Neuron 55~(6) (2007) 919--929.
	
	\bibitem{gu2002neuromodulatory}
	Q.~Gu, Neuromodulatory transmitter systems in the cortex and their role in
	cortical plasticity, Neuroscience 111~(4) (2002) 815--835.
	
	\bibitem{reynolds2002dopamine}
	J.~N. Reynolds, J.~R. Wickens, Dopamine-dependent plasticity of corticostriatal
	synapses, Neural Networks 15~(4) (2002) 507--521.
	
	\bibitem{zhang2009gain}
	J.-C. Zhang, P.-M. Lau, G.-Q. Bi, Gain in sensitivity and loss in temporal
	contrast of {STDP} by dopaminergic modulation at hippocampal synapses,
	Proceedings of the National Academy of Sciences 106~(31) (2009) 13028--13033.
	
	\bibitem{marder2012neuromodulation}
	E.~Marder, Neuromodulation of neuronal circuits: back to the future, Neuron
	76~(1) (2012) 1--11.
	
	\bibitem{nadim2014neuromodulation}
	F.~Nadim, D.~Bucher, Neuromodulation of neurons and synapses, Current Opinion
	in Neurobiology 29 (2014) 48--56.
	
	\bibitem{fremaux2016neuromodulated}
	N.~Fr{\'e}maux, W.~Gerstner, Neuromodulated spike-timing-dependent plasticity,
	and theory of three-factor learning rules, Frontiers in neural circuits 9
	(2016) 85.
	
	\bibitem{izhikevich2007solving}
	E.~M. Izhikevich, Solving the distal reward problem through linkage of {STDP}
	and dopamine signaling, Cerebral Cortex 17~(10) (2007) 2443--2452.
	
	\bibitem{pavlov2003conditioned}
	I.~P. Pavlov, G.~V. Anrep, Conditioned reflexes, Courier Corporation, 2003.
	
	\bibitem{thorndike1898review}
	E.~L. Thorndike, Review of animal intelligence: An experimental study of the
	associative processes in animals., Psychological Review.
	
	\bibitem{farries2007reinforcement}
	M.~A. Farries, A.~L. Fairhall, Reinforcement learning with modulated spike
	timing--dependent synaptic plasticity, Journal of Neurophysiology 98~(6)
	(2007) 3648--3665.
	
	\bibitem{florian2007reinforcement}
	R.~V. Florian, Reinforcement learning through modulation of
	spike-timing-dependent synaptic plasticity, Neural Computation 19~(6) (2007)
	1468--1502.
	
	\bibitem{legenstein2008learning}
	R.~Legenstein, D.~Pecevski, W.~Maass, A learning theory for reward-modulated
	spike-timing-dependent plasticity with application to biofeedback, {PLoS}
	Computational Biology 4~(10) (2008) e1000180.
	
	\bibitem{vasilaki2009spike}
	E.~Vasilaki, N.~Fr{\'e}maux, R.~Urbanczik, W.~Senn, W.~Gerstner, Spike-based
	reinforcement learning in continuous state and action space: when policy
	gradient methods fail, {PLoS} Computational Biology 5~(12) (2009) e1000586.
	
	\bibitem{fremaux2010functional}
	N.~Fr{\'e}maux, H.~Sprekeler, W.~Gerstner, Functional requirements for
	reward-modulated spike-timing-dependent plasticity, Journal of Neuroscience
	30~(40) (2010) 13326--13337.
	
	\bibitem{friedrich2011spatio}
	J.~Friedrich, R.~Urbanczik, W.~Senn, Spatio-temporal credit assignment in
	neuronal population learning, {PLoS} Computational Biology 7~(6) (2011)
	e1002092.
	
	\bibitem{fremaux2013reinforcement}
	N.~Fr{\'e}maux, H.~Sprekeler, W.~Gerstner, Reinforcement learning using a
	continuous time actor-critic framework with spiking neurons, {PLoS}
	Computational Biology 9~(4) (2013) e1003024.
	
	\bibitem{hoerzer2014emergence}
	G.~M. Hoerzer, R.~Legenstein, W.~Maass, Emergence of complex computational
	structures from chaotic neural networks through reward-modulated {Hebbian}
	learning, Cerebral Cortex 24~(3) (2014) 677--690.
	
	\bibitem{masquelier2007unsupervised}
	T.~Masquelier, S.~J. Thorpe, Unsupervised learning of visual features through
	spike timing dependent plasticity, {PLoS} Computational Biology 3~(2) (2007)
	e31.
	
	\bibitem{brader2007learning}
	J.~M. Brader, W.~Senn, S.~Fusi, Learning real-world stimuli in a neural network
	with spike-driven synaptic dynamics, Neural Computation 19~(11) (2007)
	2881--2912.
	
	\bibitem{querlioz2013immunity}
	D.~Querlioz, O.~Bichler, P.~Dollfus, C.~Gamrat, Immunity to device variations
	in a spiking neural network with memristive nanodevices, IEEE Transactions on
	Nanotechnology 12~(3) (2013) 288--295.
	
	\bibitem{yu2013rapid}
	Q.~Yu, H.~Tang, K.~C. Tan, H.~Li, Rapid feedforward computation by temporal
	encoding and learning with spiking neurons, IEEE transactions on neural
	networks and learning systems 24~(10) (2013) 1539--1552.
	
	\bibitem{lee2016training}
	J.~H. Lee, T.~Delbruck, M.~Pfeiffer, Training deep spiking neural networks
	using backpropagation, Frontiers in Neuroscience 10.
	
	\bibitem{o2016deep}
	P.~O'Connor, M.~Welling, Deep spiking networks, arXiv preprint
	arXiv:1602.08323.
	
	\bibitem{kheradpisheh2017stdp}
	S.~R. Kheradpisheh, M.~Ganjtabesh, S.~J. Thorpe, T.~Masquelier, Stdp-based
	spiking deep convolutional neural networks for object recognition, Neural
	Networks.
	
	\bibitem{thiele2017wake}
	J.~Thiele, P.~U. Diehl, M.~Cook, A wake-sleep algorithm for recurrent, spiking
	neural networks, arXiv preprint arXiv:1703.06290.
	
	\bibitem{diehl2015unsupervised}
	P.~U. Diehl, M.~Cook, Unsupervised learning of digit recognition using
	spike-timing-dependent plasticity, Frontiers in Computational Neuroscience 9
	(2015) 99.
	
	\bibitem{cao2015spiking}
	Y.~Cao, Y.~Chen, D.~Khosla, Spiking deep convolutional neural networks for
	energy-efficient object recognition, International Journal of Computer Vision
	113~(1) (2015) 54--66.
	
	\bibitem{tavanaei2016bio}
	A.~Tavanaei, A.~S. Maida, Bio-inspired spiking convolutional neural network
	using layer-wise sparse coding and {STDP} learning, arXiv preprint
	arXiv:1611.03000.
	
	\bibitem{merolla2011digital}
	P.~Merolla, J.~Arthur, F.~Akopyan, N.~Imam, R.~Manohar, D.~S. Modha, A digital
	neurosynaptic core using embedded crossbar memory with {45pJ} per spike in
	45nm, in: Custom Integrated Circuits Conference (CICC), 2011 IEEE, IEEE,
	2011, pp. 1--4.
	
	\bibitem{hussain2014improved}
	S.~Hussain, S.-C. Liu, A.~Basu, Improved margin multi-class classification
	using dendritic neurons with morphological learning, in: Circuits and Systems
	(ISCAS), 2014 IEEE International Symposium on, IEEE, 2014, pp. 2640--2643.
	
	\bibitem{oconnor2014realtime}
	P.~O'Connor, D.~Neil, S.-C. Liu, T.~Delbruck, M.~Pfeiffer, Real-time
	classification and sensor fusion with a spiking deep belief network,
	Frontiers in Neuroscience 7 (2013) 178.
	
	\bibitem{beyeler2013categorization}
	M.~Beyeler, N.~D. Dutt, J.~L. Krichmar, Categorization and decision-making in a
	neurobiologically plausible spiking network using a {STDP}-like learning
	rule, Neural Networks 48 (2013) 109--124.
	
	\bibitem{diehl2015fast}
	P.~U. Diehl, D.~Neil, J.~Binas, M.~Cook, S.-C. Liu, M.~Pfeiffer,
	Fast-classifying, high-accuracy spiking deep networks through weight and
	threshold balancing, in: Neural Networks (IJCNN), 2015 International Joint
	Conference on, IEEE, 2015, pp. 1--8.
	
	\bibitem{zhao2015feedforward}
	B.~Zhao, R.~Ding, S.~Chen, B.~Linares-Barranco, H.~Tang, Feedforward
	categorization on {AER} motion events using cortex-like features in a spiking
	neural network, IEEE Transactions on Neural Networks and Learning Systems
	26~(9) (2015) 1963--1978.
	
	\bibitem{ponulak2010supervised}
	F.~Ponulak, A.~Kasi{\'n}ski, Supervised learning in spiking neural networks
	with {ReSuMe}: sequence learning, classification, and spike shifting, Neural
	Computation 22~(2) (2010) 467--510.
	
	\bibitem{neftci2012event}
	E.~Neftci, S.~Das, B.~Pedroni, K.~Kreutz-Delgado, G.~Cauwenberghs, Event-driven
	contrastive divergence for spiking neuromorphic systems., Frontiers in
	Neuroscience 7 (2012) 272--272.
	
	\bibitem{tavanaei2016acquisition}
	A.~Tavanaei, T.~Masquelier, A.~S. Maida, Acquisition of visual features through
	probabilistic spike-timing-dependent plasticity, in: International Joint
	Conference on Neural Networks (IJCNN), IEEE, 2016, pp. 307--314.
	
	\bibitem{kheradpisheh2016bio}
	S.~R. Kheradpisheh, M.~Ganjtabesh, T.~Masquelier, Bio-inspired unsupervised
	learning of visual features leads to robust invariant object recognition,
	Neurocomputing 205 (2016) 382--392.
	
	\bibitem{thorpe1989biological}
	S.~J. Thorpe, M.~Imbert, Biological constraints on connectionist modelling,
	Connectionism in Perspective (1989) 63--92.
	
	\bibitem{vanrullen2002surfing}
	R.~VanRullen, S.~J. Thorpe, Surfing a spike wave down the ventral stream,
	Vision Research 42~(23) (2002) 2593--2615.
	
	\bibitem{krizhevsky2012imagenet}
	A.~Krizhevsky, I.~Sutskever, G.~E. Hinton, Imagenet classification with deep
	convolutional neural networks, in: Advances in Neural Information Processing
	Systems, 2012, pp. 1097--1105.
	
	\bibitem{song2000competitive}
	S.~Song, K.~D. Miller, L.~F. Abbott, Competitive {Hebbian} learning through
	spike-timing-dependent synaptic plasticity, Nature Neuroscience 3~(9) (2000)
	919--926.
	
	\bibitem{guyonneau2005neurons}
	R.~Guyonneau, R.~VanRullen, S.~J. Thorpe, Neurons tune to the earliest spikes
	through {STDP}, Neural Computation 17~(4) (2005) 859--879.
	
	\bibitem{simonyan2014very}
	K.~Simonyan, A.~Zisserman, Very deep convolutional networks for large-scale
	image recognition, arXiv preprint arXiv:1409.1556.
	
	\bibitem{thorpe1996speed}
	S.~Thorpe~J, D.~Fize, C.~Marlot, Speed of processing in the human visual
	system, Nature 381~(6582) (1996) 520.
	
	\bibitem{hung2005fast}
	C.~P. Hung, G.~Kreiman, T.~Poggio, J.~J. DiCarlo, Fast readout of object
	identity from macaque inferior temporal cortex, Science 310~(5749) (2005)
	863--866.
	
	\bibitem{dicarlo2007untangling}
	J.~J. DiCarlo, D.~D. Cox, Untangling invariant object recognition, Trends in
	Cognitive Sciences 11~(8) (2007) 333--341.
	
	\bibitem{liu2009timing}
	H.~Liu, Y.~Agam, J.~R. Madsen, G.~Kreiman, Timing, timing, timing: fast
	decoding of object information from intracranial field potentials in human
	visual cortex, Neuron 62~(2) (2009) 281--290.
	
	\bibitem{dicarlo2012does}
	J.~J. DiCarlo, D.~Zoccolan, N.~C. Rust, How does the brain solve visual object
	recognition?, Neuron 73~(3) (2012) 415--434.
	
	\bibitem{fukushima1982neocognitron}
	K.~Fukushima, S.~Miyake, Neocognitron: A self-organizing neural network model
	for a mechanism of visual pattern recognition, in: Competition and
	Cooperation in Neural Nets, Springer, 1982, pp. 267--285.
	
	\bibitem{lecun1995convolutional}
	Y.~LeCun, Y.~Bengio, Convolutional networks for images, speech, and time
	series, The Handbook of Brain Theory and Neural Networks 3361~(10) (1995)
	1995.
	
	\bibitem{serre2007robust}
	T.~Serre, L.~Wolf, S.~Bileschi, M.~Riesenhuber, T.~Poggio, Robust object
	recognition with cortex-like mechanisms, IEEE Transactions on Pattern
	Analysis and Machine Intelligence 29~(3).
	
	\bibitem{lee2009convolutional}
	H.~Lee, R.~Grosse, R.~Ranganath, A.~Y. Ng, Convolutional deep belief networks
	for scalable unsupervised learning of hierarchical representations, in:
	Proceedings of the 26th Annual International Conference on Machine Learning,
	ACM, 2009, pp. 609--616.
	
	\bibitem{zeiler2014visualizing}
	M.~D. Zeiler, R.~Fergus, Visualizing and understanding convolutional networks,
	in: European Conference on Computer Vision, Springer, 2014, pp. 818--833.
	
	\bibitem{kheradpisheh2016humans}
	S.~R. Kheradpisheh, M.~Ghodrati, M.~Ganjtabesh, T.~Masquelier, Humans and deep
	networks largely agree on which kinds of variation make object recognition
	harder, Frontiers in Computational Neuroscience 10~(74) (2016) 92.
	
	\bibitem{kheradpisheh2016deep}
	S.~R. Kheradpisheh, M.~Ghodrati, M.~Ganjtabesh, T.~Masquelier, Deep networks
	can resemble human feed-forward vision in invariant object recognition,
	Scientific Reports 6 (2016) 32672.
	
	\bibitem{furber2016large}
	S.~Furber, Large-scale neuromorphic computing systems, Journal of neural
	engineering 13~(5) (2016) 051001.
	
	\bibitem{ruckauer2017conversion}
	B.~R{\"u}ckauer, I.-A. Lungu, Y.~Hu, M.~Pfeiffer, S.-C. Liu, Conversion of
	continuous-valued deep networks to efficient event-driven networks for image
	classification, Front. Neurosci. 11: 682. doi: 10.3389/fnins.
	
	\bibitem{courbariaux2015binaryconnect}
	M.~Courbariaux, Y.~Bengio, J.-P. David, Binaryconnect: Training deep neural
	networks with binary weights during propagations, in: Advances in Neural
	Information Processing Systems, 2015, pp. 3123--3131.
	
	\bibitem{binas2016deep}
	J.~Binas, G.~Indiveri, M.~Pfeiffer, Deep counter networks for asynchronous
	event-based processing, arXiv preprint arXiv:1611.00710.
	
	\bibitem{esser2016convolutional}
	S.~K. Esser, P.~A. Merolla, J.~V. Arthur, A.~S. Cassidy, R.~Appuswamy,
	A.~Andreopoulos, D.~J. Berg, J.~L. McKinstry, T.~Melano, D.~R. Barch, et~al.,
	Convolutional networks for fast, energy-efficient neuromorphic computing,
	Proceedings of the National Academy of Sciences (2016) 201604850.
	
	\bibitem{gardner2014classifying}
	B.~Gardner, I.~Sporea, A.~Gr{\"u}ning, Classifying spike patterns by
	reward-modulated {STDP}, in: International Conference on Artificial Neural
	Networks, Springer, 2014, pp. 749--756.
	
	\bibitem{huerta2009fast}
	R.~Huerta, T.~Nowotny, Fast and robust learning by reinforcement signals:
	explorations in the insect brain, Neural Computation 21~(8) (2009)
	2123--2151.
	
	\bibitem{mnih2015human}
	V.~Mnih, K.~Kavukcuoglu, D.~Silver, A.~A. Rusu, J.~Veness, M.~G. Bellemare,
	A.~Graves, M.~Riedmiller, A.~K. Fidjeland, G.~Ostrovski, et~al., Human-level
	control through deep reinforcement learning, Nature 518~(7540) (2015)
	529--533.
	
	\bibitem{silver2016mastering}
	D.~Silver, A.~Huang, C.~J. Maddison, A.~Guez, L.~Sifre, G.~Van Den~Driessche,
	J.~Schrittwieser, I.~Antonoglou, V.~Panneershelvam, M.~Lanctot, et~al.,
	Mastering the game of {Go} with deep neural networks and tree search, Nature
	529~(7587) (2016) 484--489.
	
\end{thebibliography}

\newpage
\clearpage

\normalsize



\section*{Supplementary material}

# "Supplementary Materials"
# First-spike based visual categorization using reward-modulated STDP

Milad Mozafari[1,2], Saeed Reza Kheradpisheh[1,2], Timothée Masquelier[3], Abbas Nowzari-Dalini[1], and Mohammad Ganjtabesh[1,2,*]

[1] *Department of Computer Science, School of Mathematics, Statistics, and Computer Science, University of Tehran, Tehran, Iran*

[2] *School of Biological Sciences, Institute for Research in Fundamental Sciences (IPM) , Tehran, Iran*

[3] *CerCo UMR 5549, CNRS Université Toulouse 3, France*

## 1 Datasets

In this section, a detailed description of the datasets that are used in our experiments is provided.

In all of the benchmarks, images in the training set were used to train the network and extract discriminative features, while those in the testing set were used to evaluate the network. In other words, in each training iteration, the network receives all of the images in the training set and does per-image synaptic plasticity, after which it goes under evaluation by receiving images from the testing set. This way, we assessed the ability of the network to extract general features, that work for categorization of either the training (seen) or the testing (unseen) samples. For each dataset, an exhaustive search was performed to find suitable values for all of the parameters. We should add that the size of the input windows were chosen according to the average size of the target objects in each dataset. Table SI summarizes the parameters for each dataset.

Caltech face/motorbike was chosen as the first benchmark to monitor the behavior of our network in a two-class object recognition task. Images in this task fall into two categories: faces and motorbikes (see Fig. S1a), that are provided by the California Institute of Technology[1]. For each task, 435 images were randomly selected from each category, among which 200 images were chosen for training while the remaining were used for testing. For computational efficiency, images are scaled by the factor 0.5.

The second benchmark that we chose for object recognition is ETH-80, provided by Max Planck Institute for Informatics[2]. This dataset contains eight categories of objects, that are apple, car, cow, cup, dog, horse, pear, and tomato (see Fig. S1b). In each category, there are 10 different instances, for each, 41 images are captured from different view points. Hence, this benchmark is a good assessment of view-invariant object recognition ability. Here, we applied the leave-one-out cross-validation technique to evaluate the network's performance. In each evaluation trial, nine instances of each ob-

---

[*]Corresponding author.
Email addresses:
milad.mozafari@ut.ac.ir (MM),
kheradpisheh@ut.ac.ir (SRK),
timothee.masquelier@cnrs.fr (TM),
nowzari@ut.ac.ir (AND),
mgtabesh@ut.ac.ir (MG).

[1]http://www.vision.caltech.edu
[2]https://www.mpi-inf.mpg.de/departments/computer-vision-and-multimodal-computing

1

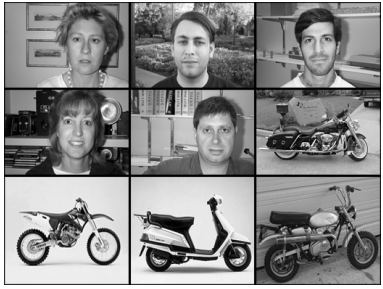

(a)

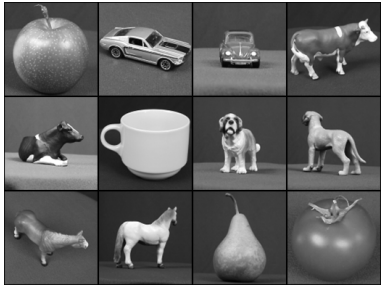

(b)

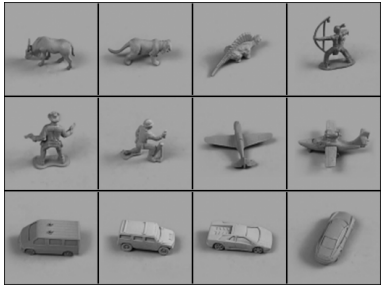

(c)

Figure S1: Sample images from the three datasets that are used in this work. (a) Caltech face/motorbike, (b) ETH-80, and (c) NORB dataset.

ject were put in the training set ($8 \times 9 \times 41 = 2952$ in total) and the remaining ones constituted the testing set ($8 \times 41 = 328$).

The third object recognition benchmark is called NORB, provided by Courant Institute at New York University[3]. This dataset contains five classes of objects, that are four-legged animals, human figures, airplanes, trucks, and cars (see Fig. S1c). The object instances (50 different object in total) are imaged in different view-points, coupled with different lighting conditions. In this task, we used the training and testing sets provided by the authors of this dataset. Since there are large number of images in this dataset, we also performed the

---

[3] http://www.cs.nyu.edu/~ylclab/data/norb-v1.0/

following sub-sampling procedure on both training and testing sets. This sub-sampling results in images of only one camera (left), one lighting condition (level 4), four elevations (odd values), and all of the 18 azimuths. Therefore, there are 3600 ($50 \times 4 \times 18$) images that are equally divided into two 1800-image sets: training and testing samples.

## 2 Results

In this part of the supplementary materials, we present some complementary explanations and demonstrations on the results of the experiments.

### 2.1 Object Recognition

#### 2.1.1 Feature Extraction

Using R-STDP, the proposed network is able to extract discriminative features. This is of great importance for similar objects like apples and tomatoes. Fig. S2 clearly shows how the network distinguishes apples from tomatoes by including tomato's head in the corresponding feature.

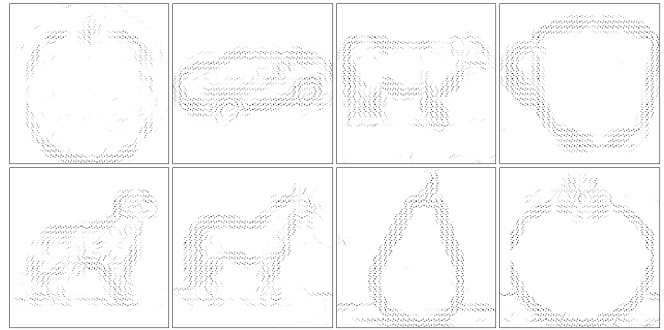

Figure S2: Some of the features extracted for each category of the ETH-80 dataset.

#### 2.1.2 Confusion Matrix

As STDP tends to find frequent features, it encounters problems in distinguishing similar objects. The reason is that neurons learn common features and start to fire for images in two or more categories. Regarding this tendency, the external classifier has

2

Table SI: Values for parameters used in each of the object recognition tasks.

| Dataset | $\omega_{c1}$ | $\omega_{s2}$ | $n$ | $\mathcal{T}$ | $a_r^+$ | $a_r^-$ | $a_p^+$ | $a_p^-$ | $p_{drop}$ |
|---|---|---|---|---|---|---|---|---|---|
| Caltech | 7 | 17 | 20 | 42 | 0.005 | −0.0025 | 0.0005 | −0.005 | 0.5 |
| ETH-80 | 5 | 31 | 80 | 160 | 0.01 | −0.0035 | 0.0006 | −0.01 | 0.4 |
| NORB | 5 | 23 | 50 | 150 | 0.05 | −0.003 | 0.0005 | −0.05 | 0.5 |

no other means than getting biased towards one of the similar categories.

On the other hand, R-STDP helps the network to extract features that provide more balanced (not biased) discrimination. Comparing the confusion matrices of R-STDP and STDP, you can see that the former has a more symmetric structure than the latter. To provide a quantified comparison, for each matrix $A$, we computed the following equation:

$$asym(A) = \frac{\sum_i \sum_j |A_{ij} - A_{ji}|}{2}, \quad (1)$$

where $A_{ij}$ denotes the value of row $i$ and column $j$. $asym$ can be considered as a measure of being asymmetry, where as the matrix $A$ is more symmetry, the value of $asym(A)$ is closer to 0 (see Fig. S3).

## 2.2 Dropout

One of the techniques we used to avoid overfitting is dropout. In each training iteration, dropout turns some neurons off by chance. This procedure temporarily decreases the network's complexity and forces smaller portion of neurons to cover the whole training set as much as possible. This way, the probability that a neuron only learns few samples decreases, and the total involvement rate of neurons in solving the task increases. Fig. S4 illustrates the involvement rate of neurons for one of the trials with each of the ETH-80 and NORB datasets. Green bars denote the number of times that a neuron generates earliest spike for testing images belonging to its pre-assigned category, while red bars denote the number of false alarms. By comparing this measure with and without dropout, we found that dropout improves performance by increasing the chance of finding more discriminative features, as well as decreasing the rate of blind firings.

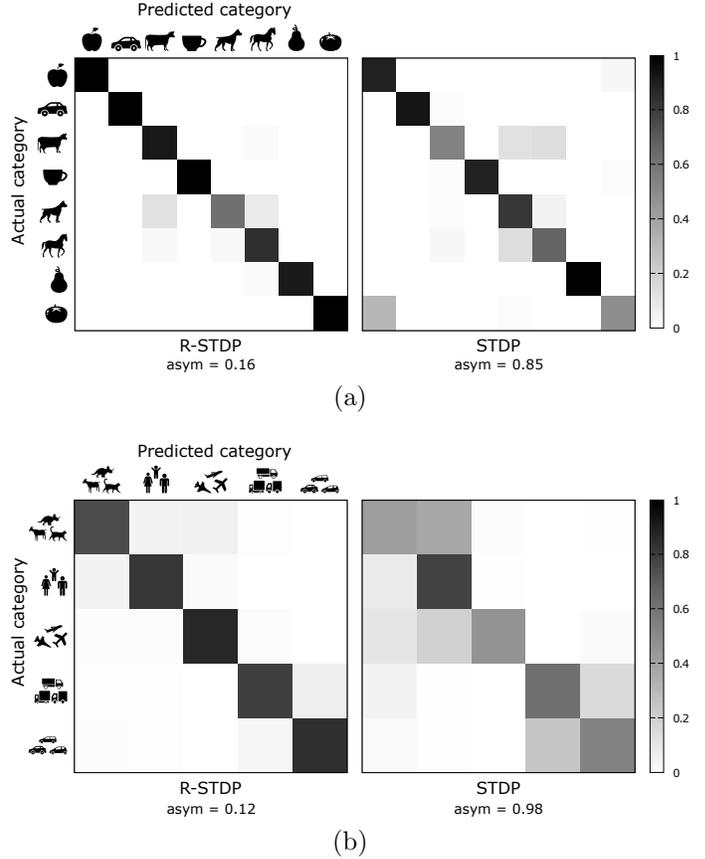

Figure S3: Comparing confusion matrices of the network over the (a) ETH-80, and (b) NORB datasets, while employing R-STDP and STDP.

## 2.3 Shallow CNN

In object recognition tasks, we compared the proposed model to a shallow CNN with a similar structure. Here you can find the detailed structure of this CNN.

Input images to this CNN are grayscale. The first convolutional layer has four feature maps with the same input window size as our Gabor filters in layer $S1$. This layer is followed by a max-pooling layer with the same pooling window size and stride as the neurons in layer $C1$. Next, we put a convolutional layer with $n$ feature maps and neurons

3

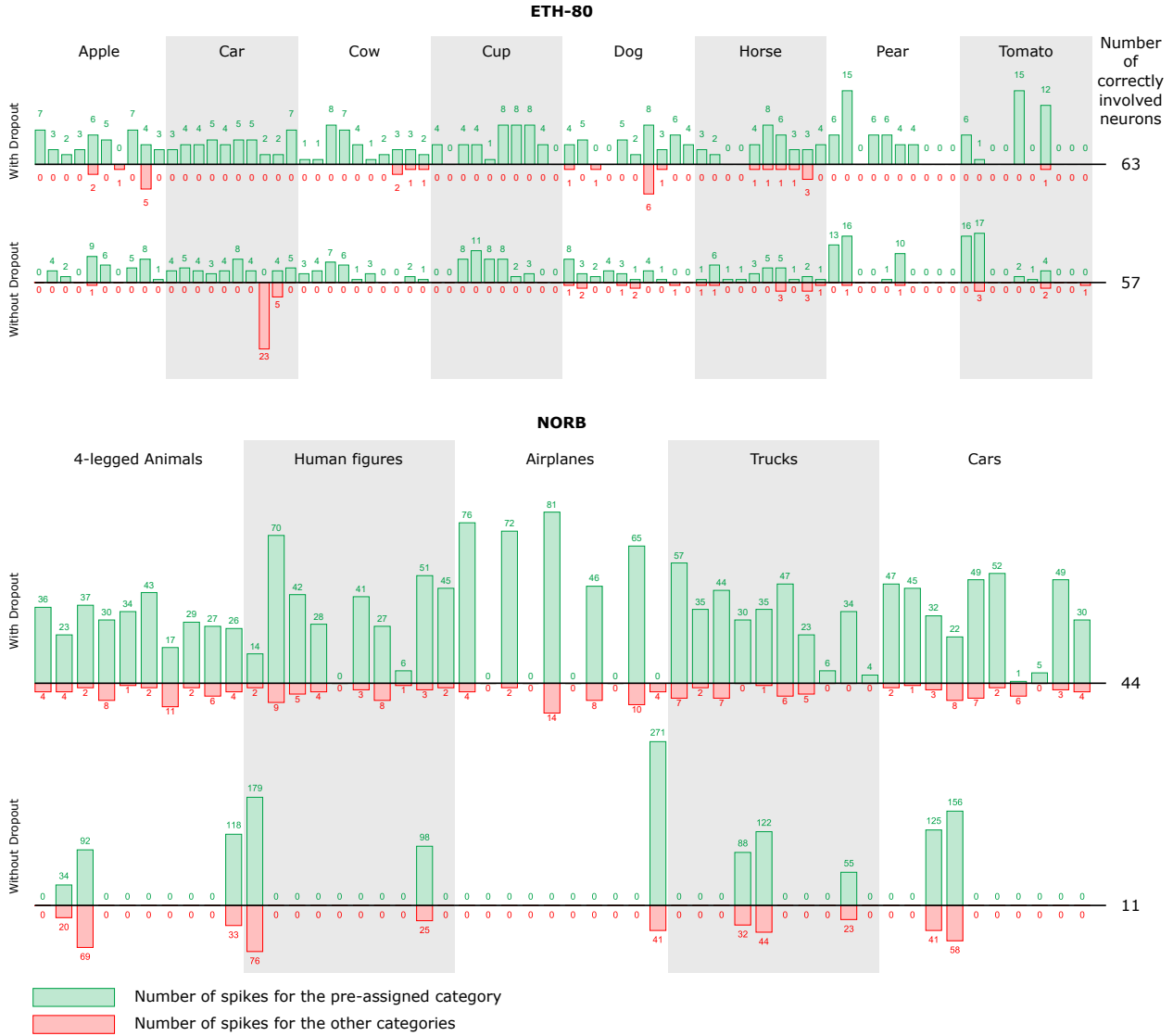

Figure S4: Involvement rate of neurons in solving different object recognition tasks with and without applying dropout.

Table SII: Performance of VGG16 over the ETH-80 and NORB datasets.

| Dataset | Fine-tuned | Trained from Scratch |
|---------|------------|----------------------|
| ETH-80  | 97.8       | 96.8                 |
| NORB    | 94.7       | 80.1                 |

with input window that covers the whole incoming data from previous layer. Activation functions of convolutional neurons are set to ReLU.

After convolutional layers, we put two dense layers: a hidden and an output layer. Number of neurons in the hidden layer is tuned for each task. Activation function for the hidden and output layer are set to ReLU and Softmax, respectively.

For overfitting avoidance, we explore using dropout on dense layers and kernel regularizers with different parameter values to achieve the best performance.

## 2.4 Comparison with Deep Convolutional Neural Network

Deep convolutional neural networks (DCNNs) has achieved outstanding results in various complex visual object recognition tasks. Spiking neural net-

works, despite of being theoretically more powerful, cannot beat them in practice by the time of writing this manuscript. One problem with DCNNs is that they are data-hungry and they need lots of training samples to generalize well. This problem can be tackled by applying fine-tuning strategies over pre-trained DCNNs (initialized with pre-trained weights), instead of training them from scratch (initialized with random weights).

Here, we repeated our object recognition tasks, this time using a DCNN called VGG16, which is known as one of the state-of-the-art networks for object categorization. We used Keras network API[4] with Tensorflow[5] as its backend to examine VGG16 network, while doing fine-tuning or training from scratch. Table SII summarizes the best performances of VGG16 over the ETH-80 and NORB (after doing the aforementioned subsampling) datasets. Since the performance depends on the fine-tuning strategy and values of parameters, we do not guaranty that the obtained results are the best possible ones. However, we did our best to achieve high performances. According to the results, VGG16 always beats us when it is fine-tuned, whereas it has encountered serious overfitting problem, as well as the proposed network, while being trained on the NORB dataset from scratch.

---

[4]https://keras.io/
[5]https://www.tensorflow.org/



\end{document}